\documentclass[5p, times]{elsarticle}

\usepackage{amssymb}
\usepackage{amsthm}
\usepackage{xspace}
\usepackage{ecrc}
\usepackage{color}
\usepackage{bm}
\usepackage{url}
\usepackage{ulem}

\journal{Nuclear Instruments and Methods in Physics Research, section A}
\volume{00}
\firstpage{1}

\begin{document}
\begin{frontmatter}

\title{
High-energy extension of the gamma-ray band observable with an electron-tracking Compton camera
}

\cortext[cor]{Corresponding author}
\author[R1,R2]{Tomohiko~Oka\corref{cor}}
\ead{toka@fc.ritsumei.ac.jp}
\author[Kyoto]{Shingo~Ogio}
\author[Kyoto]{Mitsuru~Abe}
\author[UMBC]{Kenji~Hamaguchi}
\author[Kyoto]{Tomonori~Ikeda}
\author[Kyoto]{Hidetoshi~Kubo}
\author[Tohoku]{Shunsuke~Kurosawa}
\author[Kobe]{Kentaro~Miuchi}
\author[ISAS]{Yoshitaka~Mizumura}
\author[Kyoto]{Yuta~Nakamura}
\author[Kanazawa]{Tatsuya~Sawano}
\author[Kyoto]{Atsushi~Takada}
\author[Kyoto]{Taito~Takemura}
\author[Kyoto2]{Toru~Tanimori}
\author[Kyoto]{Kei~Yoshikawa}

\affiliation[R1]{
    organization = {Department of Physical Sciences, Ritsumeikan University},
    addressline = {1-1-1 NojiHigashi}, 
    city = {Kusatsu}, 
    postcode = {525-8577},
    state = {Shiga},
    country = {Japan}
}

\affiliation[R2]{
    organization = {Research Organization of Science and Technology, Ritsumeikan University},
    addressline = {1-1-1 NojiHigashi}, 
    city = {Kusatsu}, 
    postcode = {525-8577},
    state = {Shiga},
    country = {Japan}
}

\affiliation[Kyoto]{
    organization = {Graduate School of Science, Kyoto University},
    addressline  = {Kitashirakawa-Oiwakecho},
    city         = {Kyoto},
    postcode     = {606-8502},
    state        = {Kyoto},
    country      = {Japan}
}

\affiliation[UMBC]{
    organization = {Department of Physics, University of Maryland},
    addressline  = {Baltimore County 1000 Hilltop Circle},
    city         = {Baltimore},
    postcode     = {21250},
    state        = {Maryland},
    country      = {USA}
}

\affiliation[Tohoku]{
    organization = {Institute of Materials Research, Tohoku University},
    addressline  = {Katahira 2-1-1, Aoba},
    city         = {Sendai},
    postcode     = {980-8577},
    state        = {Miyagi},
    country      = {Japan}
}

\affiliation[Kobe]{
    organization = {Graduate School of Science, Kobe University},
    addressline  = {1-1 Rokkodai, Nada},
    city         = {Kobe},
    postcode     = {657-8501},
    state        = {Hyogo},
    country      = {Japan}
}

\affiliation[ISAS]{
    organization = {Institute of Space and Astronautical Science, Japan Aerospace Exploration Agency},
    addressline  = {Yoshinodai 3-1-1, Chuou},
    city         = {Sagamihara},
    postcode     = {252-5210},
    state        = {Kanagawa},
    country      = {Japan}
}

\affiliation[Kanazawa]{
    organization = {Graduate School of Natural Science and Technology, Kanazawa University},
    addressline  = {Kakuma},
    city         = {Kanazawa},
    postcode     = {920-1192},
    state        = {Ishikawa},
    country      = {Japan}
}

\affiliation[Kyoto2]{
    organization = {Institute for Integrated Radiation and Nuclear Science, Kyoto University},
    addressline = {2, Asashiro-Nishi, Kumatori-cho},
    city = {Sennan-gun},
    postcode = {590-0494},
    state = {Osaka},
    country = {Japan}
}

\begin{abstract}

Although the MeV gamma-ray band is a promising energy-band window in astrophysics, the current situation of MeV gamma-ray astronomy  significantly lags  behind those of the other energy bands in angular resolution and sensitivity.
An electron-tracking Compton camera (ETCC), a next-generation MeV detector, is expected to  revolutionize the situation. 
An ETCC tracks each Compton-recoil electron with a gaseous electron tracker and  determines the incoming direction of each gamma-ray photon; thus, it has a strong background rejection power and yields a better angular resolution than classical Compton cameras.  
Here, we study ETCC events in which the Compton-recoil electrons do not deposit all energies to the electron tracker but escape and hit the surrounding pixel scintillator array (PSA). 
The PSA provides additional information on the electron-recoil direction, which enables us to improve significantly  the angular resolution. 
We developed an analysis method for this untapped class of events and applied it to laboratory and simulation data. 
We found that the energy spectrum obtained from the simulation agreed with that of the actual data within a factor of 1.2.
We then evaluated the detector performance using the simulation data.
The angular resolution for the new-class events was found to be twice as good as in the previous study at the energy range 1.0--2.0~MeV, where both analyses overlap.
We also found that the total effective area is dominated by the contribution of the double-hit events above an energy of 1.5~MeV.
Notably, applying this new method extends the sensitive energy range with the ETCC from 0.2--2.1 MeV in the previous studies to up to 3.5~MeV.
Adjusting the PSA dynamic range should improve the sensitivity in even higher energy gamma-rays.
The development of this new analysis method would pave the way for future observations by ETCC to fill the MeV-band sensitivity gap in astronomy.
\end{abstract}

\begin{keyword}
Gamma detectors \sep Time projection chambers \sep Gaseous detectors \sep Scintillators \sep Compton imaging
\end{keyword}

\end{frontmatter}


\section{Introduction} \label{sec:introduction}

MeV gamma-ray observations are expected to be a key to solving many puzzles in astrophysics~\cite[see, e.g.,][for  reviews and references therein]{Volker2001, Pinkau2009, Hamaguchi2019, DeAngelis2021}.
Observations of de-excitation nuclear-line emission at 4--6~MeV \cite{Benhabiles2013ApJ} and the characteristic cutoff structure of the $\pi^{0}$-decay spectrum at ${\sim} 70$~MeV \cite{Dermer1986A&A} would potentially provide conclusive evidence for the origin of cosmic rays (CRs).
The spatial distribution of the $^{26}$Al nuclear line at 1.8~MeV tells us where nucleosynthesis occurs in our Galaxy~\cite{Knodlseder1999ApJ}.
Moreover, the Compton telescope COMPTEL onboard the Compton Gamma-Ray Observatory satellite, a pioneer in MeV gamma-ray astronomy,  discovered an anomalously bright emission toward the Galactic Center at 1--30~MeV~\cite{Strong1996A&AS}.
This so-called ``COMPTEL excess'' cannot be explained by contributions from known Galactic sources and suggests the presence of radiation from dark matter annihilation/decay \cite{Boddy2015PRD}, primordial black holes \cite{Carr2010PRD}, or unknown sources \cite{Tsuji2023ApJ}.
However, the sensitivity and angular resolution of none of the past MeV gamma-ray telescopes have been good enough to resolve the above-mentioned scientific targets observable at $>1$~MeV.

\begin{figure}
    \centering
    \includegraphics[width=\linewidth]{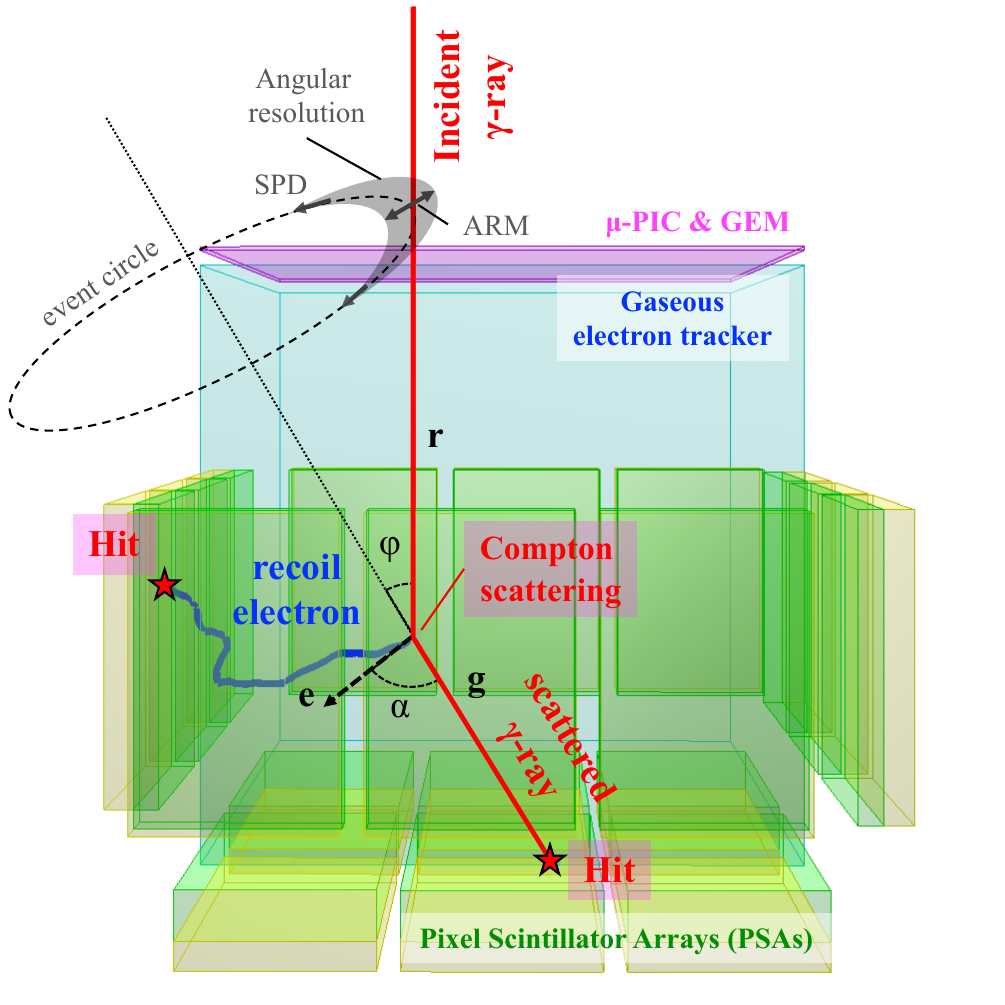}
    \caption{
    Schematic view of  a double-hit event for the ETCC.
    Red and blue lines show the paths of gamma rays and electrons, respectively. 
    The red stars represent the hit points at the PSAs.
    As with  Compton-scattered gamma rays, the recoil electron reaches the PSA through the TPC.
    Parameters related to event reconstruction ($\mathbf{r}$, $\mathbf{g}$, $\mathbf{e}$, $\phi$, $\alpha$) and PSF (ARM, SPD) are also shown in this figure.
    }
    \label{fig:Concept}
\end{figure}

A major difficulty in MeV-band observations is that conventional Compton cameras (e.g., COMPTEL \cite{Schoenfelder1993ApJS} and COSI \cite{Tomsick2019BAAS}) cannot uniquely determine the arrival directions of gamma rays, resulting in observation data significantly contaminated by diffuse or isotropic background \cite{Schonfelder2004NewAR}.
For each incoming gamma-ray, the conventional Compton camera can only measure the direction of a scattered gamma-ray from its incident position and the scattering angle from its deposited energy; as a result, it can only constrain the incoming gamma-ray direction within an annulus region in the sky.
More specifically, the angular resolution (i.e., the point spread function; PSF) of a Compton camera depends on the angular resolution measure (ARM) and scatter plane deviation (SPD) \citep[e.g.,][]{Kierans2022}.
The ARM of a Compton event is defined as the angular distance between the known source location and  the Compton event circle. 
The SPD is the angle between the true gamma-ray direction and the estimated one along the Compton event circle.
Fig.~\ref{fig:Concept} illustrates the ARM and SPD.
Since conventional Compton cameras measure the direction of scattered gamma rays but not recoil electrons, they cannot constrain the SPD.
The resultant practical angular resolution of conventional Compton cameras is worse than 20$^{\circ}$ in half-power radius (HPR) for 662~keV, even when the ARM is as good as $2^{\circ}$~\cite{Tanimori2015ApJ}\footnote{Note that the ARM is sometimes referred to as the ``angular resolution'' of a conventional Compton camera, although it does not correspond to the true PSF size.}.

To overcome these problems, we have been developing an electron-tracking Compton camera (ETCC), which is one of the next-generation MeV gamma-ray cameras and consists of a gaseous electron tracker as the Compton scattering target and pixel scintillator arrays (PSAs) as the absorbers \cite{Tanimori2004NewAR}.
The gaseous electron tracker on the ETCC obtains the three-dimensional tracks and energies of the Compton-recoil electrons generated by an incoming gamma-ray photon, whereas the PSAs measure the absorption points and energies of the scattered gamma rays.
Utilizing the complete information (direction and energies) of the recoil electrons and the scattered gamma rays, the ETCC constrains both the ARM and SPD and can uniquely determine the arrival directions of incident gamma rays. 
In other words, the ETCC establishes a one-to-one correspondence (bijection) between the incident and reconstructed gamma ray  (see \cite{Takada2022ApJ, Bernard2022} for more details).
Consequently, the ETCC has a better angular resolution and stronger background-rejection power than conventional gamma-ray cameras, achieving an unprecedented sensitivity in the MeV gamma-ray band \cite{Tanimori2015ApJ, Tanimori2017NatSR}.
In recent years, the SMILE-2+ experiment  succeeded in detecting an astronomical source, the Crab Nebula, with a balloon-borne ETCC in the energy band 0.15--2.1~MeV, even though the observation time was limited to 5.1~hours \cite{Takada2022ApJ}.
The detection is a milestone in the history of MeV gamma-ray observations, confirming the estimate of the sensitivity computed from the PSF and the effective area (Fig.~20 in \cite{Takada2022ApJ}).

So far, ETCCs are sensitive for energies up to 2~MeV, not sufficient to cover the energy band in which the aforementioned scientific targets are observable.
There are two reasons for this: first, the cross section of Compton scattering in a gas decreases with energy, and second, recoil electrons in high-energy events are more likely to escape from the TPC because of their high energy.
Among the two points, the second point can be, in principle, circumvented to a large extent through reconstruction of the Compton scattering event if the PSA is arranged to absorb the escaped electrons. 
Fig.~\ref{fig:Concept} shows schematics of the detection of a type of event that we name a ``double-hit'' event, of which the recoil electron is absorbed in the PSA.
Previous studies~\cite[e.g.,][]{Takada2011ApJ, Takada2022ApJ} did not  utilize  double-hit events for simplicity and only used ``single-hit'' events where the gas detector completely absorbs a recoil electron; the PSA of the detector did not detect signals from recoil electrons.
In this work, we analyze double-hit events for high-energy extensions of the ETCC energy band.
By utilizing double-hit events, we expect to improve not only the effective area for high-energy gamma rays but also the angular resolution.
The precision in determining the recoil direction in such a way is fundamentally better than that of single-hit events because recoil electrons in double-hit events have relatively high energies, suppressing the effects of multiple scattering in the TPC gas \cite{Bernard2022}.
The endpoint (and start point) of the recoil-electron trajectory of a double-hit event can be identified geometrically using hit information at PSAs (further details will be provided in Sect.~\ref{sec:rec}), whereas for single-hit events, this is estimated with the skewness of the track image \cite{Dujmic2007NIMA} and  time-over-threshold information \cite{Tanimori2015ApJ}, the method of which involves a significant amount of uncertainty \cite{Ikeda2021PTEP}.
As a result, the method based on double-hit events improves the accuracy of the recoil direction, leading to better SPDs in double-hit events than those obtained with the original analysis with single-hit events applied for the SMILE-2+ experiment.
We note that analyzing pair production events may help to attain a large effective area, but the angular resolution based on pair production events at the MeV gamma-ray band of ${\lesssim}10$~MeV is fundamentally worse than that based on Compton scattering events~\cite[e.g.,][]{Ueno2011NIMA}.
For this reason, we focus here only on Compton scattering events.

In this work, we establish an event reconstruction method for double-hit events and apply the method to laboratory and simulation data.
We furthermore evaluate the detector performance using the simulation data.
The details of the ETCC and the event reconstruction method are described in Sect.~\ref{sec:instrument} and Sect.~\ref{sec:rec}, respectively.
We present the analysis results and performance studies using laboratory and simulation data in Sect.~\ref{sec:results} and discuss these results in Sect.~\ref{sec:discussion}.
The summary is given in Sect.~\ref{sec:summary}.

\section{Instruments} \label{sec:instrument}

We use the ETCC detector onboard the SMILE-2+ experiment~\cite{Takada2022ApJ}, which has a sensitive volume of $30 \times 30 \times 30~\rm{cm^{3}}$.
The gaseous electron tracker in the ETCC employs a time-projection chamber (TPC) consisting of a micro pixel chamber \cite[$\mu$-PIC;][]{Ochi2001NIMPA, Miuchi2003ITNS} and a gas electron multiplier \cite[GEM;][]{Sauli1997NIMPA, Tamagawa2006NIMA} with a  100-$\mathrm{\mu m}$ liquid crystal polymer insulator.
We combine two $\mu$-PIC readout strips laid out side by side to save power consumption; the readout pitch is $800~\rm{\mu m}$.
We filled the TPC with an argon-based gas at 2 atm. Its  pressure ratio was $\rm{ Ar:CF_{4}:iso C_{4}H_{10}} =95:3:2$, which was  the same as in the T2K experiment \cite{Karlen2010NIMA}.
The energy resolution of the tracker is $45.9\%$ in full width at half maximum (FWHM) at 0.043~MeV ($\rm{GdK\alpha}$).
The PSAs are made of GSO ($\rm{Gd_{2}SiO_{5}:Ce}$) and are installed to cover the bottom and half of the side planes of the electron tracker.
Each array has $8 \times 8$ pixels with a pixel size of $6\times6~\rm{mm^{2}}$.
The scintillators are 26 (13) mm thick at the bottom (side) plane of the electron tracker, the value of which corresponds to two (one) radiation lengths.
Each module also has multi-anode photo-multiplier tubes (Hamamatsu Photonics, flat-panel H8500) to detect photons from the scintillators.
For readout, the so-called four-channel charge division method with a resistor network \cite{Sekiya2006NIMPA} is employed.
The energy resolutions of the bottom and side PSAs at 0.662~MeV are $13.4\%$ and $10.9\%$ in FWHM, respectively.

It should be noted that the PSA is covered with Enhanced Specular Reflector (ESR) films (3M~Company) to collect light, Teflon tapes to support the module, and black tapes to block light originating from electron avalanches and gas ionization in the TPC detector.
The thicknesses (and densities)  are 0.0065~cm ($1.38~\rm{g\, cm^{-3}}$) for the ESR film, 0.01~cm ($2.2~\rm{g\,cm^{-3}}$) for the Teflon tape, and 0.022~cm ($1.35~\rm{g\,cm^{-3}}$) for the black tape, respectively.
The minimum undetected energy loss for electrons that traverse the ``dead'' zone vertically, is ${\sim}100$~keV, which we need to take into account.

The SMILE-2+ ETCC issues a DAQ trigger with coincident signals of the PSA and TPC --- a TPC signal within 10-$\rm{\mu s}$ from a rising edge of a PSA wave signal~\cite{Mizumoto2015NIMA}. 
We tagged events with signals from the two PSAs as double-hit events. 

\section{Event reconstruction} \label{sec:rec}

In order to reconstruct a Compton scattering event, we first find a scattering point from ETCC data and then estimate the directions of recoil electrons and scattered gamma rays.
Conventionally, a Compton scattering point is identified with the so-called time-over-threshold skewness method \cite[e.g.,][]{Ikeda2021PTEP}, in which the head and tail of the recoil electrons are determined from the skewness of the track image \cite{Dujmic2007NIMA}, and the recoil direction is determined from time-over-threshold information \cite{Tanimori2015ApJ}.
In our method, we can geometrically identify the Compton scattering point for double-hit events. 
In a double-hit event, the hit at one of the two PSAs originates from scattered gamma rays, which does not leave any trace along the trajectory in the TPC from the scattering point. 
The other hit originates from the recoil electron that escaped from the TPC, which leaves a trace of the trajectory from the scattering point to the end of the sensitive area in the TPC with ionization energy loss. 
Since the dead zone between the TPC's sensitive area and the PSA is narrow, we can safely assume that the PSA hit by the recoil electron is closer to the endpoint of the TPC trajectory. 
Specifically, we first search for the longest trajectory in the TPC data, which should be the trajectory of the recoil electron. 
Then, we pair the endpoint with the closer PSA hit point. 
We flag the other PSA hit as the one which originates from the scattered gamma-ray.
Finally, the direction of the incident gamma ray and its energy are estimated in the procedures described in Sect.~\ref{sec:rec:dir} and Sect.~\ref{sec:rec:ene}, respectively, on the basis of the geometry of the Compton-scattering event.

\subsection{Direction of incident gamma rays}
\label{sec:rec:dir}

The method to estimate the arrival directions of gamma rays is the same as in single-hit event analyses~\cite[e.g.,][]{Takada2022ApJ} and is concisely described below.
The unit vector of the incident gamma-ray, $\mathbf{r}$, is obtained as:
\begin{equation}
    \mathbf{r} = \left( \rm{cos}\,\phi - \frac{\rm{sin}\,\phi}{\rm{tan}\,\alpha} \right) \mathbf{g} + \left( \frac{\rm{sin}\,\phi}{\rm{sin} \,\alpha} \right) \mathbf{e}
\end{equation}
where $\mathbf{g}$ and $\mathbf{e}$ are unit vectors in the directions of the scattered gamma-ray and the recoil electron, respectively, in the laboratory system, $\alpha$ is the angle between $\mathbf{g}$ and $\mathbf{e}$, and $\phi$ is the scattering angle.
The vector $\mathbf{g}$ is obtained by linearly interpolating the Compton scattering point and the PSA hit point,  whereas $\mathbf{e}$ is obtained by fitting the TPC data within 2~cm from the Compton scattering point with a Hesse function.
This fit range was chosen by considering a balance between the effects of multiple scattering and data statistics.
The angle $\phi$ is given with the energies of the recoil electron ($E_{\rm{e}}$ ) and the scattered gamma ray ($E_{\gamma}$) by:
\begin{equation}
   \textrm{cos}\,\phi = 1 - m_{\rm{e}} c^{2} \left( \frac{1}{E_{\gamma}} - \frac{1}{E_{\gamma} + E_{\rm{e}}} \right). 
\end{equation}

\subsection{Energy estimation}
\label{sec:rec:ene}

The incident gamma-ray energy is estimated according to the following form:
\begin{equation}
\label{eq:totalenergy}
    E = E_{\rm{e},\,\rm{TPC}} + E_{\rm{e},\,\rm{DZ}} +  E_{\rm{e},\,\rm{PSA}} + E_{\gamma,\,\rm{PSA}},
\end{equation}
where the first subscript indicates the particle species (electron (e) or $\gamma$-ray photon) and the second indicates the location of energy loss (in the\ TPC, PSA, or DZ, dead zone). 
The quantities $E_{\rm{TPC}}$ and $E_{\rm{PSA}}$ are directly obtained from the measurement at each detector.  However, we need to consider a different way to estimate $E_{\rm{e},\,\rm{DZ}}$.
If we assume that the recoil electron goes straight through the dead zone with the incident angle  $\theta_{\rm{ter}}$, its energy  is given by:
\begin{equation}
    \label{eq:E_e_dz}
    E_{\rm{e,\,DZ}} = \frac{1}{\rm{cos}\,\theta_{\rm{ter}}} \sum_{i = \rm{PVC,~PTFE, ~PET}} \rho_{i}d_{i} \times  f_{i}(E_{\rm{e},\,\rm{PSA}}),
\end{equation}
where $\rho$, $d$, and $f$ are the density, thickness, and stopping power of each matter in the dead zone, respectively, and
 PVC, PTFE, and PET indicate materials of the ESR film, the Teflon tape, and the black tape, respectively.
The angle $\theta_{\rm{ter}}$ is obtained by fitting the TPC data within 2~cm from the end point (in a similar way to $\mathbf{e}$, but not the start point).

\subsection{Event selection}
\label{sec:rec:cut}

Below are the criteria for the gamma-ray event selection.
The criteria are similar to the single-hit event analysis \cite[e.g.,][]{Takada2022ApJ}, but we made some changes and additions for the double-hit event analysis.

\begin{itemize}
    \item {\bf Energy Cut:} 
    The energy criteria are mainly given according to the dynamic range of the detector. 
    Although the original dynamic range of each pixel in the PSA is different from each other, the band covered with those of $>90\%$ of pixels is defined as the effective dynamic range of the PSA, which is 0.15--2.1~MeV \cite{Takada2022ApJ}.
    We thus give the criteria on the energies in the PSAs for both the scattered gamma ray and the recoil electron: $0.15~\rm{MeV} < E_{PSA} < 2.1~\rm{MeV}$.
    We also give an additional cut, $E_{\rm{e,\,PSA}} > 400~\rm{keV}$, given that the uncertainty of the energy deposit estimate is significant below this energy range.
    We discuss the issue in detail in Sect.~\ref{sec:lab}.
    
    \item {\bf dE/dx:} 
    The correlation between the energy loss of a particle in a electron tracker and its track length ($L$)  renders the particle identification into the following three types: i) the fully contained electron in the TPC (i.e., the single-hit event), ii) the minimum ionizing particles, e.g., the recoil electron that escaped from the TPC (i.e., the double-hit event) or cosmic-ray (CR) muons, and iii) other charged particles.
    We here give our criteria based on $dE/dx$ of electrons in argon gas~\cite{Sauli1977}:
    \begin{eqnarray}
    \label{eq:dedx1}
    L &>& 7.1 \left( \frac{\rho}{1~\rm{g\,cm^{-3}}} \right)^{-1} \left( \frac{E_{\rm e,\,TPC}}{1.0~\rm{MeV}} \right)^{1.5} + 7~\rm{mm}, \\
    \label{eq:dedx2}
    L &<& 270~\rm{mm}, 
    \end{eqnarray}
    where $\rho$ is the average density of the gas in the TPC.
    The criteria in Eq.~\ref{eq:dedx1} and Eq.~\ref{eq:dedx2} reject single-hit events (i) and track events by charged particles (iii), respectively.
    These criteria are similar to the previous work for single-hit event analysis (Fig.~4 in \cite{Takada2022ApJ}) but with the inequality sign in Eq.~\ref{eq:dedx1} reversed.

    \item {\bf Fiducial volume}:
    Most track data of charged particles start near the edge of the TPC, as they are likely to react with materials outside the TPC (e.g., $\mu$-PIC) before entering it.
    We therefore adopt a fiducial volume cut at the minimum depth of the track from the top of the TPC ($z_{\rm min}$) and the distance between the TPC edges and the Compton scattering point ($L_{\rm scat}$):
    \begin{eqnarray}
        &&32~\rm{mm} < z_{\rm min} < 332~\rm{mm}, \\
        &&L_{\rm{scat}} > 30~\rm{mm}.
    \end{eqnarray}
    Here, whereas the first condition is set according to the physical parameters of the ETCC, the second condition is set with  examination of the balance between the statistics of Compton gamma-ray events and the rejection efficiency of the background cosmic-ray (BGCR) events. 
    In the examination, we calculate the energy spectra of the BGCRs (which include protons, electrons, positrons, and neutrons), using the PARMA packages \cite{Sato2008RadR}, for the environment during the SMILE-2+ balloon experiment \cite{Takada2022ApJ}, which has a cutoff rigidity of 8.6~GV and is at an altitude of $3.0~\rm{g cm^{-2}}$.
    
    \item {\bf Compton-scattering kinematics test:}   
    As described in Sect.~\ref{sec:rec:dir}, the angle between the scattered gamma ray and the recoil electron, $\alpha$, is defined as:
    \begin{equation}
    \label{eq:alpha_geo}
    \rm{cos}\,\alpha_{\rm geo} = \mathbf{g} \cdot \mathbf{e}.
    \end{equation} 
    The angle $\alpha$ can be also expressed as, based on Compton-scattering kinematics:
    \begin{equation}
        \rm{cos}\,\alpha_{ \rm{kin} } = \left( 1 - \frac{{\it m}_{\rm e}c^{2}}{\it E_{\gamma}} \right) \sqrt{ \frac{E_{\rm e}} {E_{\rm e} + 2m_{\rm e}c^{2}} }.
    \end{equation}
    In the case of Compton-scattering events, this estimate should be consistent with that derived with Eq.~\ref{eq:alpha_geo}.
    We thus select the events with the condition:
    \begin{equation}
      |\rm{cos}\,\alpha_{\rm geo} - \rm{cos}\,\alpha_{\rm kin}| < 0.5.
    \end{equation}
    This criterion is the same as in the previous study~(Eq. 6 in \cite{Takada2022ApJ}).

    \item {\bf Connection of the electron track:}
    We also use the information on the connectivity of recoil-electron trajectory from the TPC to PSA.
    We adopt the minimum condition: $\theta_{\rm ter} < 90^{\circ}$.

    \item {\bf Zenith angle:}
    The earlier studies suggest that the ETCC is less efficient in detecting  gamma-ray photons coming from the detector sides and measured parameters are less accurate in such cases \cite{Takada2022ApJ}. 
    We thus only use the events with a zenith angle of $< 60^{\circ}$ to suppress background contamination and/or uncertainty.
\end{itemize}
 
We simulated the BGCR events produced by protons, electrons, positrons, and neutrons, adopting the fluxes estimated with PARMA and assuming the same environment as that during the SMILE-2+ balloon experiment.
We found that the above-described criteria eliminate $99.98\%$ of the BGCRs.

\section{Analysis results} \label{sec:results}

\subsection{Application to laboratory data}
\label{sec:lab}

\begin{figure}
    \centering
    \includegraphics[width=\linewidth]{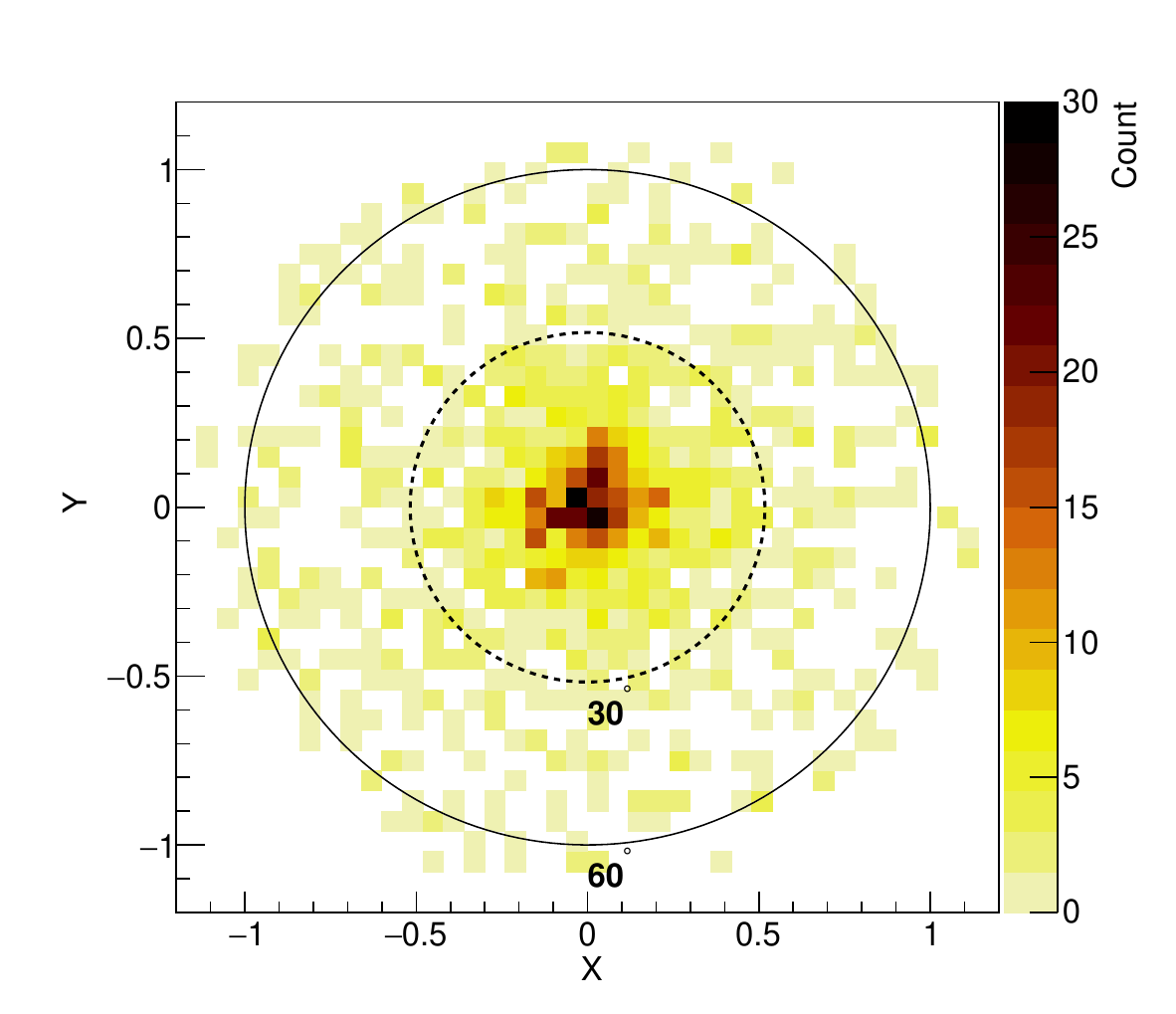}
    \caption{
    Reconstructed count map of $^{60}$Co from the laboratory data.
    The $^{60}$Co source is located  at the center of the image (X = 0, Y = 0).
    The source is placed 1.6~m away from the detector, and its angular size (viewed from the detector) is $0.1^{\circ}$.
    The dashed and solid circles show the positions corresponding to angular distances of 30$^{\circ}$ and 60$^{\circ}$, respectively.
    }
    \label{fig:60Co_map}
\end{figure}

We first apply the analysis method described in Sect.~\ref{sec:rec} to the data taken in our laboratory.
As for the radioisotope (RI) source, we use a $^{60}$Co source ($\sim 0.39$~MBq), which mainly emits two species of gamma rays with energies of 1.17 and 1.33~MeV.
The RI source is installed at a position 1.6~m away from the ETCC.
Fig.~\ref{fig:60Co_map} shows the reconstructed image of $^{60}$Co.
An emission is clearly visible at the center of the image, which agrees with the actual location of the RI source.

\begin{figure}
    \centering
    \includegraphics[width=\linewidth]{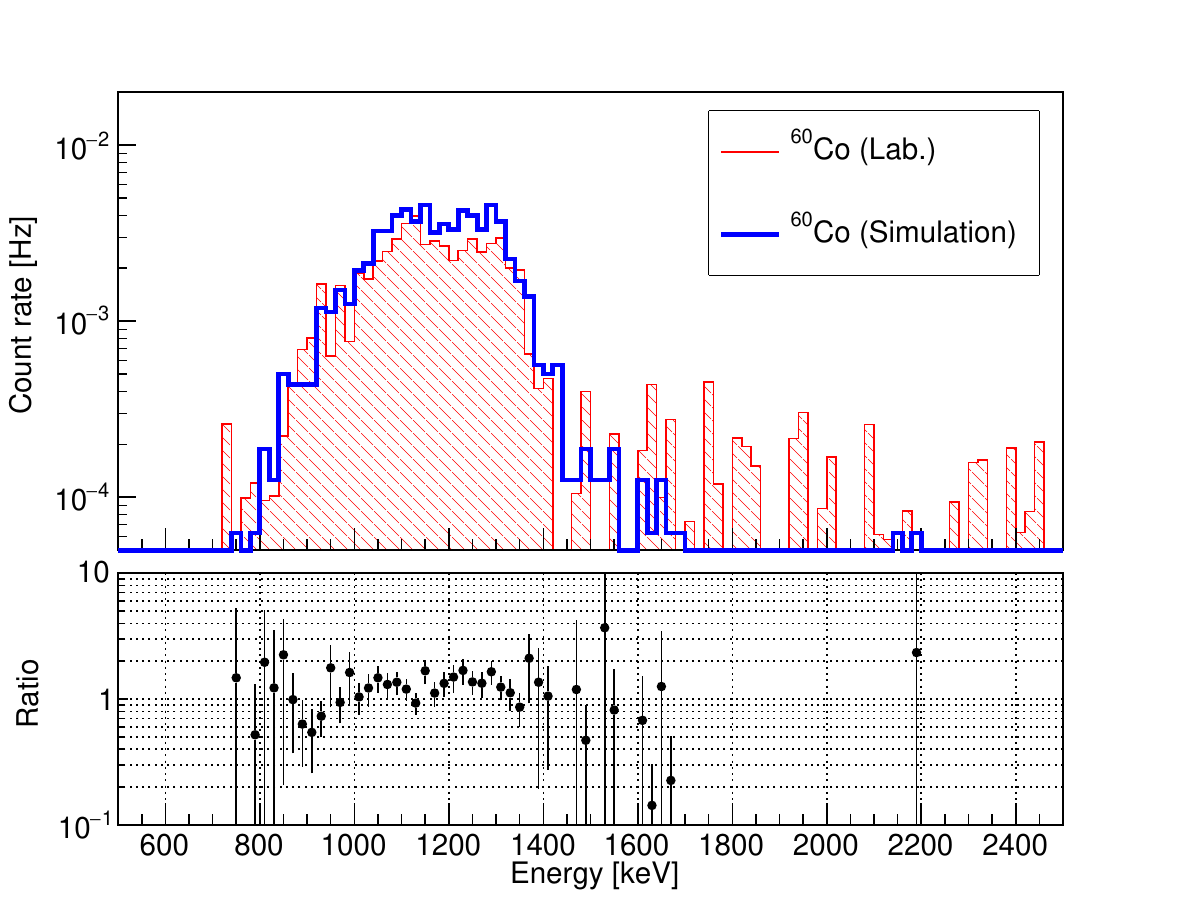}
    \caption{
    Energy spectra of $^{60}$Co (which mainly emits gamma rays of 1173 and 1333~keV) measured with the ETCC.
    In the top panel, the red histogram  shows the laboratory data,  and the blue thick line  does the simulation data.
    The bottom panel shows the ratio between the  two data.
    }
    \label{fig:60Co_spectrum}
\end{figure}

We then extract an energy spectrum from the $^{60}$Co data and show it in Fig.~\ref{fig:60Co_spectrum}.
In  this data analysis, we subtract the measurements without the RI source from those with the RI source to reduce the contamination of environmental radiation in the laboratory.
We also generate the simulation data with the simulator of SMILE-2+ ETCC~\cite{Takada2022ApJ}, the design of which is based on Geant4~\cite[version 10.04-patch02;][]{Agostinelli2002NIMA}, and also show it in Fig.~\ref{fig:60Co_spectrum}. 
We find that the simulation result is consistent with the laboratory-data result within a factor of 1.2.

\begin{figure*}
    \begin{tabular}{cc}
        \begin{minipage}{0.5\hsize}
            \centering
            \includegraphics[width=\hsize]{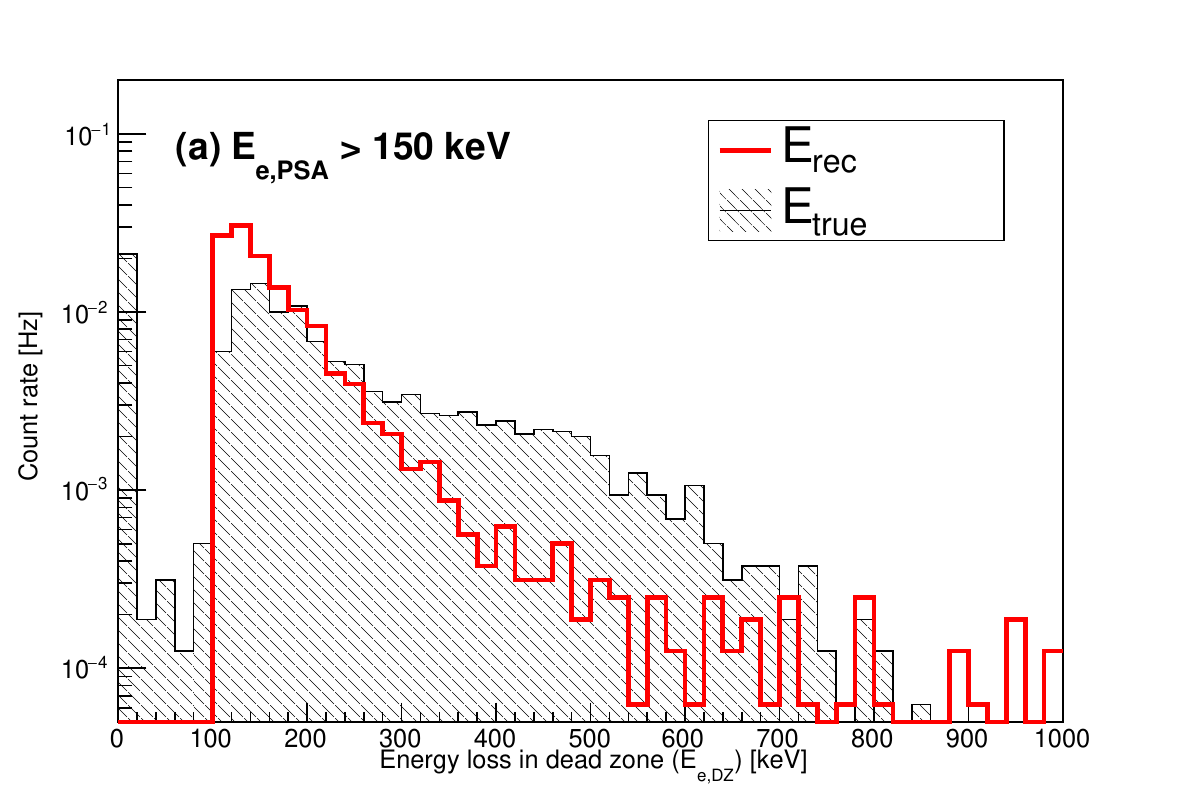}
        \end{minipage}
        &
        \begin{minipage}{0.5\hsize}
            \centering
            \includegraphics[width=\hsize]{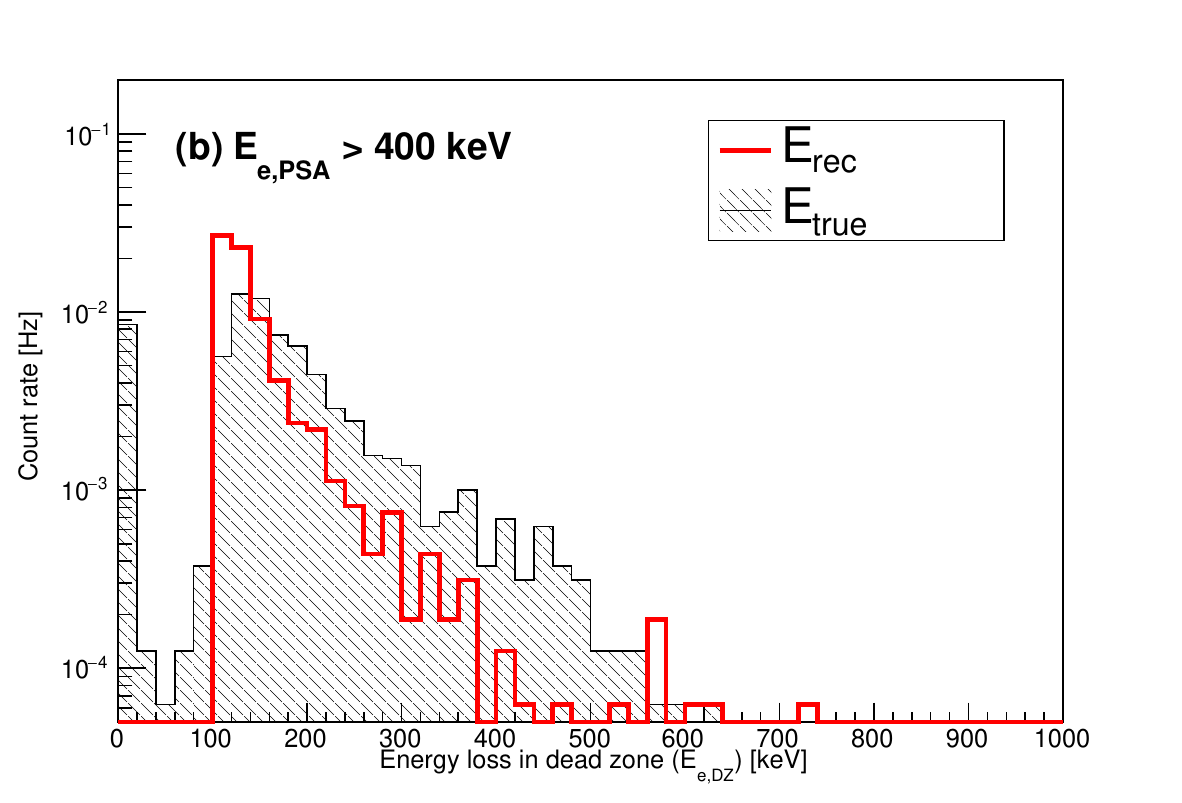}
        \end{minipage}
        \\
        \begin{minipage}{0.5\hsize}
            \centering
            \includegraphics[width=\hsize]{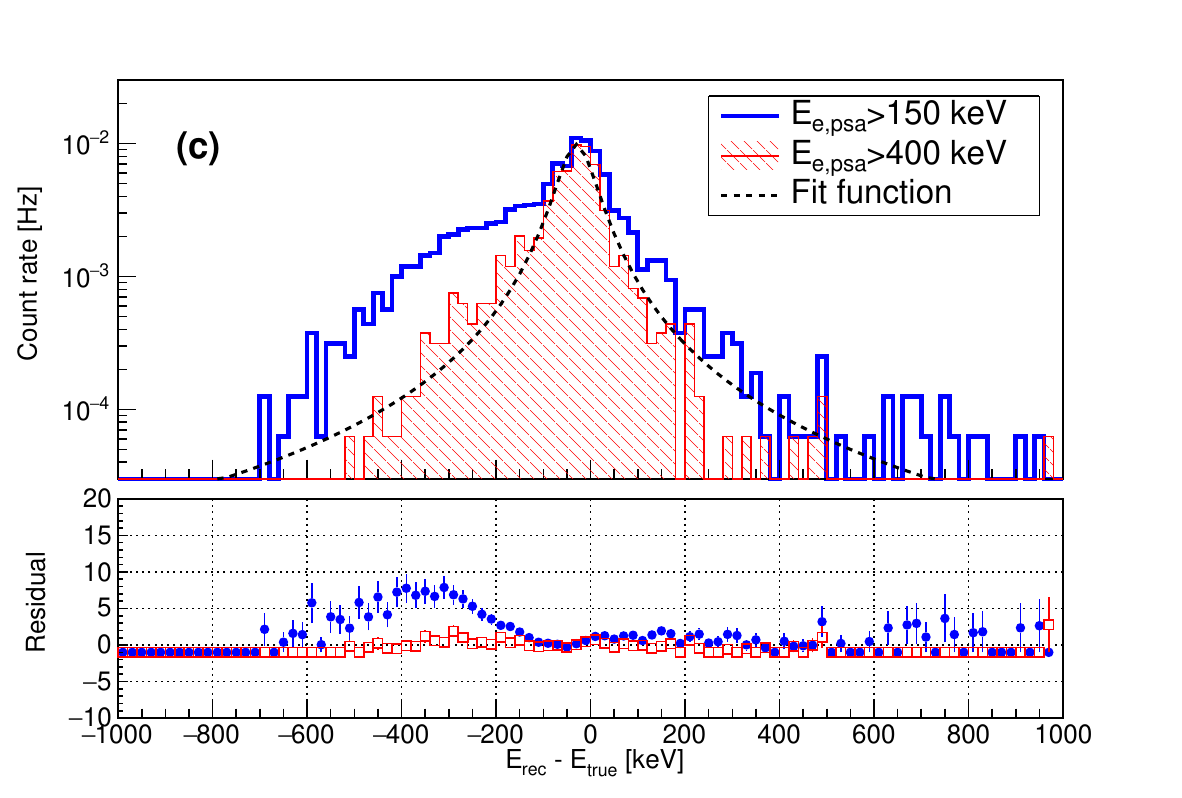}
        \end{minipage}
        &
        \begin{minipage}{0.425\hsize}
            \centering
            \includegraphics[width=\hsize]{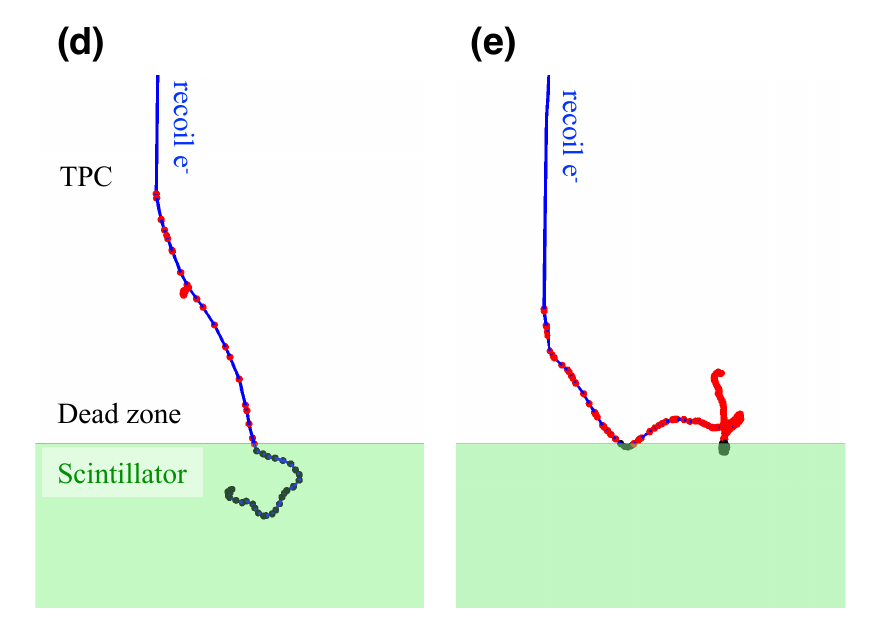}
        \end{minipage}
  \end{tabular}
    \caption{
    Obtained distributions of the energy deposit in the dead zone, $E_{\rm{e,\,DZ}}$, from simulation data.
    {\bf (a,\ b)} Black and red histograms  show the distributions of the true energy ($E_{\rm true}$) obtained in the simulations and the reconstructed energy ($E_{\rm rec}$) with the method described in Sect.~\ref{sec:rec}.
    The energy thresholds on $E_{\rm{e,\,PSA}}$  are (a) 150~keV and (b) 400~keV. 
    {\bf (c)} Discrepancy between $E_{\rm true}$ and $E_{\rm rec}$.
    Blue and red histograms show the results of $E_{\rm{e,\,PSA}} > 150~\rm{keV}$ and $> 400~\rm{keV}$, respectively.
    The red line indicates the best-fit result for the histogram for $E_{\rm{e,\,PSA}} > 400~\rm{keV}$ with a Lorentzian function.
    The lower panel displays the residual from the best-fit function, $(N_{\rm data}-N_{\rm fit})/N_{\rm fit}$.
    The meaning of colors is the same as in the upper panel.
    {\bf (d,\ e)} Examples of the simulated recoil electron trajectory around the dead zone.
    Red and black markers show the step points at the dead zone and the scintillator, respectively.
    In Panel d, the recoil electron  reaches the PSA and is absorbed there, whereas in Panel e, the electron is scattered in the PSA and returns to the dead zone.
    }
    \label{fig:DeadZone}
\end{figure*}

Fig.~\ref{fig:DeadZone} compares the reconstructed electron energy ($E_{\rm{rec}}$) at the dead zone with its true energy obtained from the simulation data ($E_{\rm{true}}$).
A small fraction of events have a true energy of $E_{\rm true} = 0~\rm{keV}$ (Panels a and b).
Examining the track paths of such events in the simulation data, we find that most of them can be  categorized into two types: (i) actual double-hit events where the recoil electrons enter PSAs from a thin part of the dead zone and (ii) originally single-hit events where the scattered gamma rays are scattered again in the PSAs and reach another scintillator. 
In case (i), the reconstructed energy with the current method is overestimated,  whereas in case (ii), we mistake  them as  double-hit events.
The fraction of the events with $E_{\rm true} = 0$~keV in all events that remain after applying the event cut described in Sect.~\ref{sec:rec:cut} is $11.6\%$, which we regard as the systematic uncertainty on flux normalization.
We note that the current simulator calculates the tracks with a coarser step size around the dead zone  than in the sensitive area of the TPC to reduce computation time.
Although this variable step-size somewhat increases the probability of case (i),  we find that the effect is not critical for our evaluation of the detector performance. 
More detailed simulations will be conducted in the future.

In addition, the $E_{\rm true}$ distribution with $E_{\rm{e,\,PSA}} > 150~\rm{keV}$ shown in Panel a has a high-energy tail at $\sim$ 500~keV, whereas the reconstructed energy ($E_{\rm rec}$)  distribution displays no such feature.
This high-energy tail does not appear for events with $E_{\rm{e,\,PSA}} > 400~\rm{keV}$ (Panel b).
Panel c shows the distribution of the discrepancy between $E_{\rm true}$ and $E_{\rm rec}$ for the events of $E_{\rm true} \neq 0$~keV.
We fit the distribution with $E_{\rm{e,\,PSA}} > 400~\rm{keV}$ with a Lorentzian function because $E_{\rm e,\,DZ}$ is a function of two variables, $E_{\rm e,\,PSA}$ and $1/\rm{cos}\,\theta_{\rm{ter}}$ (Eq.~\ref{eq:E_e_dz}), and is thus expected to follow a Lorentzian form.
The residual from the best-fit function (Panel c) shows an apparent excess only for the events with $E_{\rm{e,\,PSA}} > 150~\rm{keV}$.
We investigate the trajectories of recoil electrons to determine potential differences in the trajectories between the electrons that constitute the high-energy tail of the $E_{\rm true}$ distribution and the other electrons.
As a result, we find that most of the recoil electrons reaching the PSA are absorbed by the scintillator, as Panel d demonstrates an example trajectory.  
However, in some cases, they return to the dead zone due to multi-scattering effects (see an example trajectory in Panel e).
The latter causes the high-energy tail of $E_{\rm true}$.
To eliminate these events, we have introduced a cut condition of $E_{\rm{e,\,PSA}} > 400~\rm{keV}$    in this analysis as mentioned in Sect.~\ref{sec:rec:cut}.
The systematic uncertainty in $E_{\rm e,\,DZ}$ (Panel c) is $82.8 \pm 4.4$~keV  in FWHM, which is $6.6 \pm 0.4\%$ of the average energy of the incident gamma rays of $^{60}$Co.
This fluctuation is smaller than the energy resolution of the whole detector (as we will show in Sect.~\ref{sec:sim}) and thus does not significantly affect the analysis results.

\subsection{Performance study} \label{sec:sim}

\begin{figure}
    \centering
    \begin{tabular}{c}
        \begin{minipage}{\hsize}
            \centering
            \includegraphics[width=\hsize]{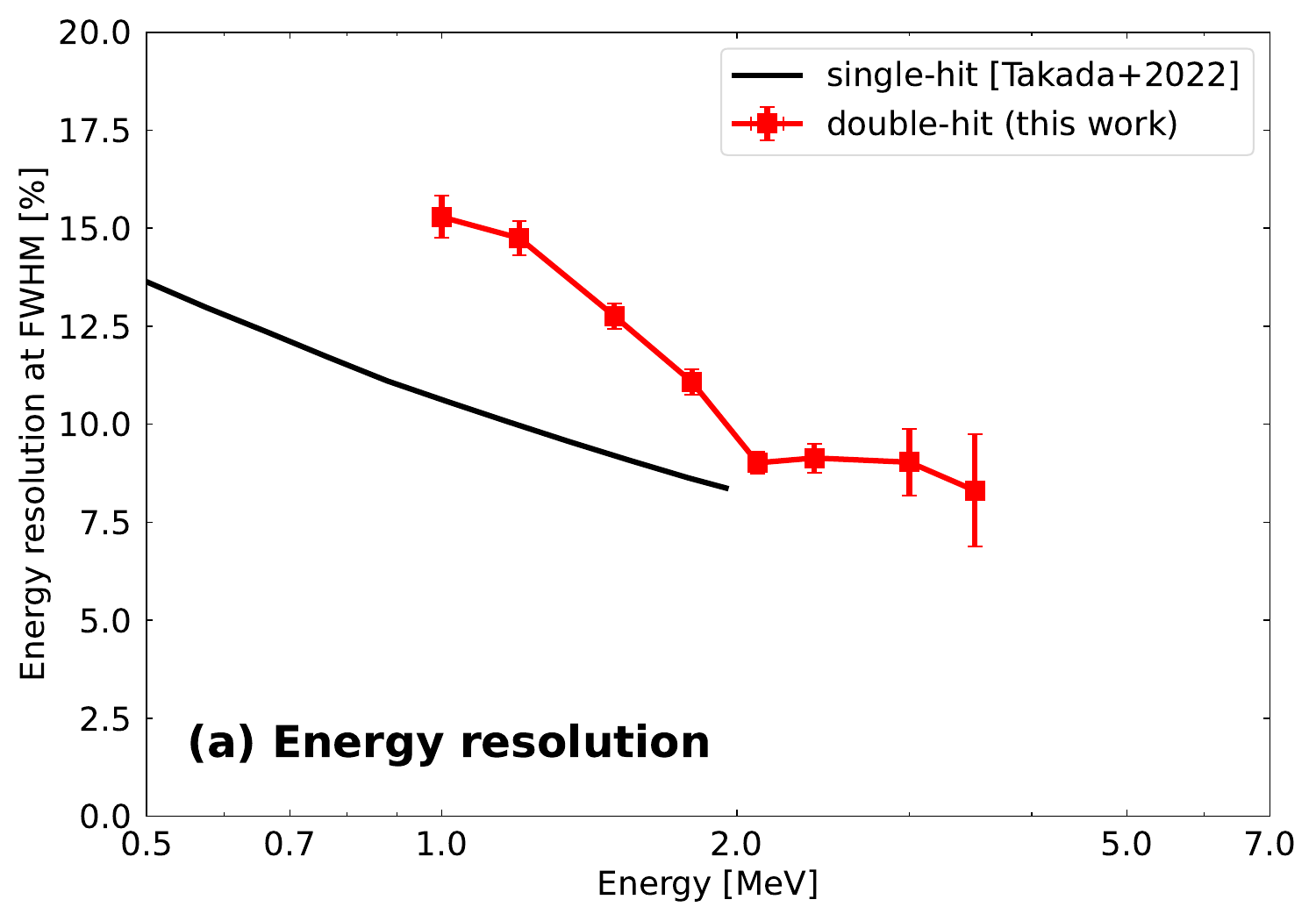}
        \end{minipage}
        \\
        \begin{minipage}{\hsize}
            \centering
            \includegraphics[width=\hsize]{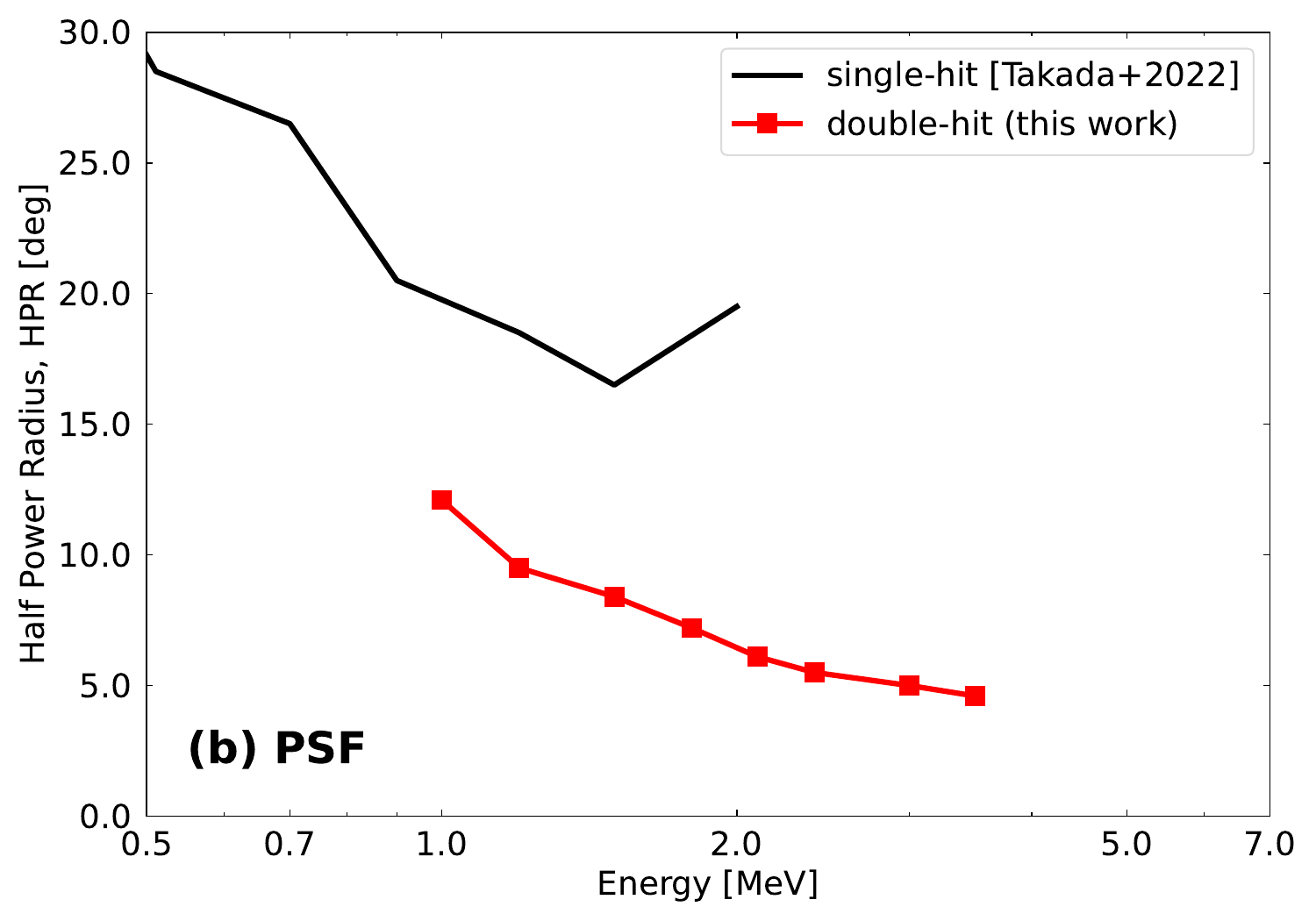}
        \end{minipage}
        \\
        \begin{minipage}{\hsize}
            \centering
            \includegraphics[width=\hsize]{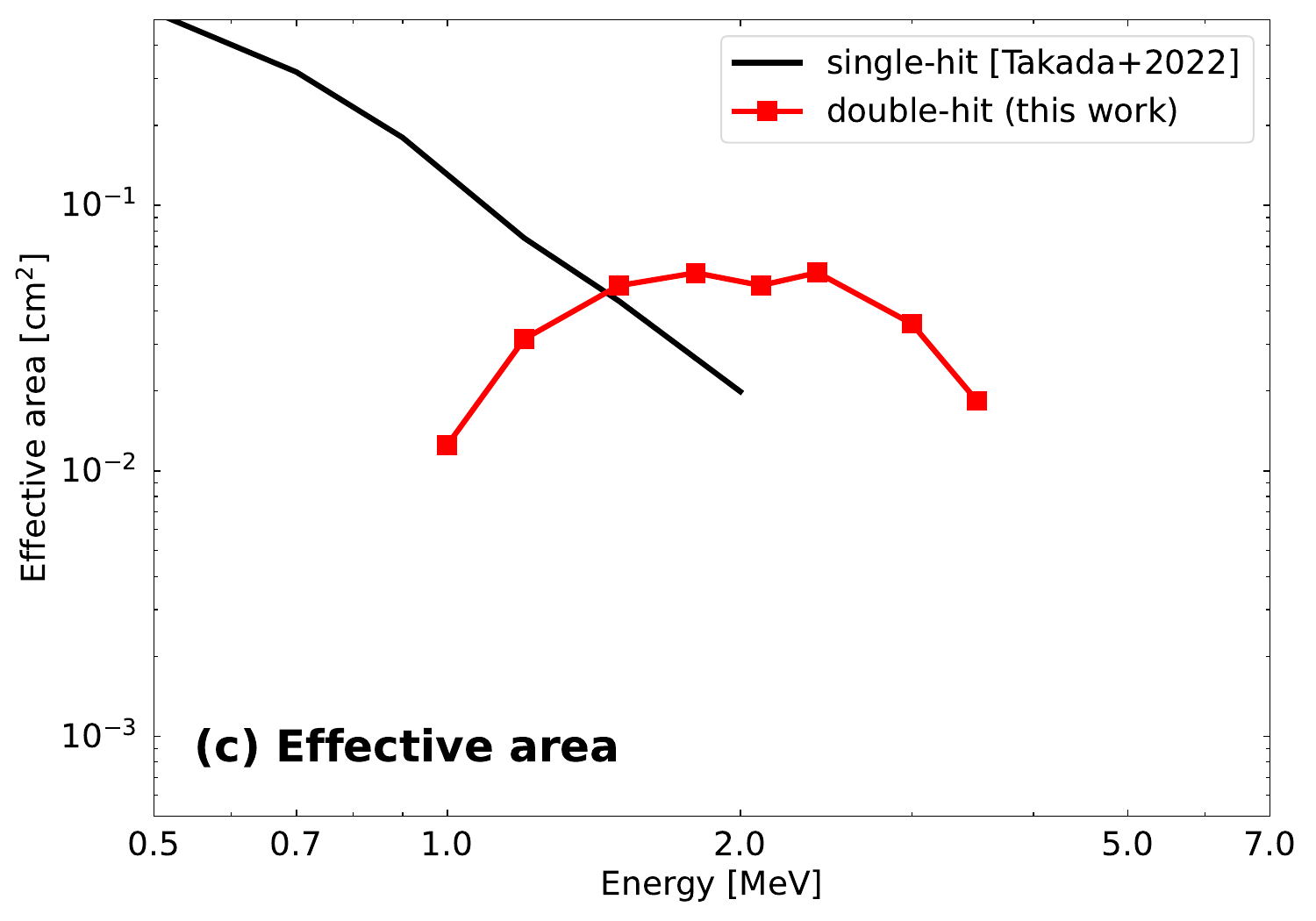}
        \end{minipage}
  \end{tabular}
    \caption{
    Results of the double-hit analysis for simulation data:  (\textbf{a}) energy resolution, (\textbf{b}) PSF, and (\textbf{c}) effective area of the ETCC,  all as a function of  incident gamma-ray energy.
    In each panel,  red markers  denote the results of double-hit events,  whereas the black line  does that of single-hit events~\citep{Takada2022ApJ}.
    }
    \label{fig:performance}
\end{figure}

We evaluate the detector performance (energy resolution, PSF, and effective area) of the double-hit event analysis at the energy range of 1.0--3.5~MeV.
For this, we conduct simulations with parallel light where monochromatic rays enter the detector at a common angle (here, a zenith angle of $0^{\circ}$).
With the simulation data, we evaluate the energy resolution, PSF, and effective area of the ETCC as a function of energy, and show them in Fig.~\ref{fig:performance}.
Here, for the PSF, we calculate the half power radius (HPR) as its representative scale.
In the evaluation of the HPR and effective area (Panels b and c), we select events whose measured energy is within twice the energy resolution in FWHM at the incident gamma-ray energy.

\begin{figure*}
    \begin{tabular}{cc}
        \begin{minipage}{0.485\hsize}
            \includegraphics[width=\hsize]{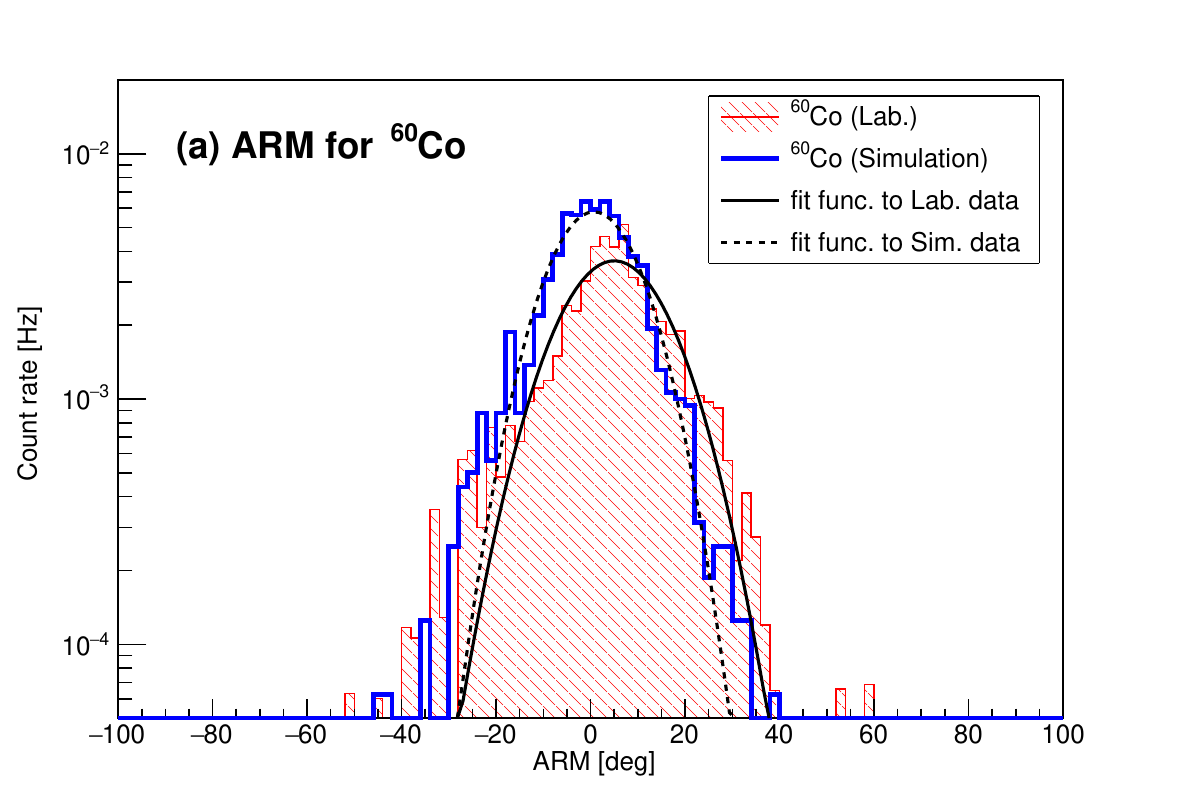}
        \end{minipage}
        &
        \hspace{-0.05\columnwidth}
        \begin{minipage}{0.485\hsize}
            \includegraphics[width=\hsize]{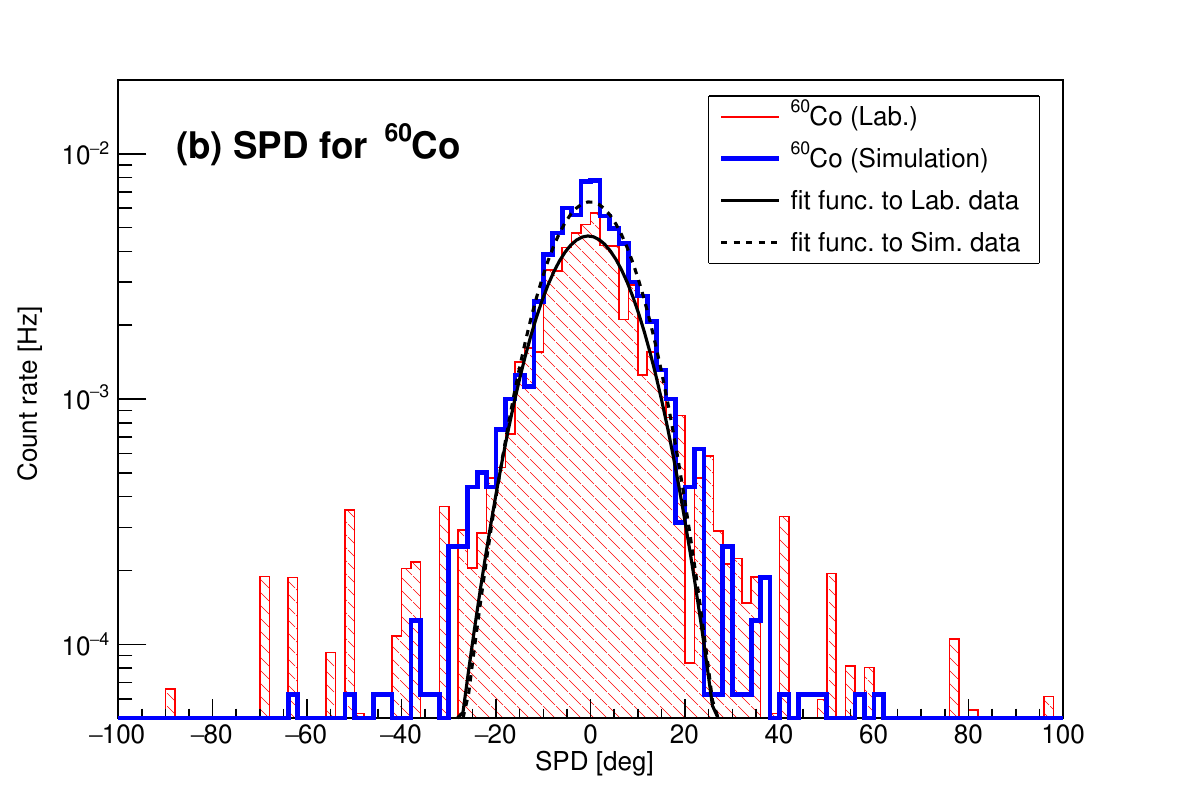}
        \end{minipage}
        \\
        \begin{minipage}{0.485\hsize}
            \includegraphics[width=\hsize]{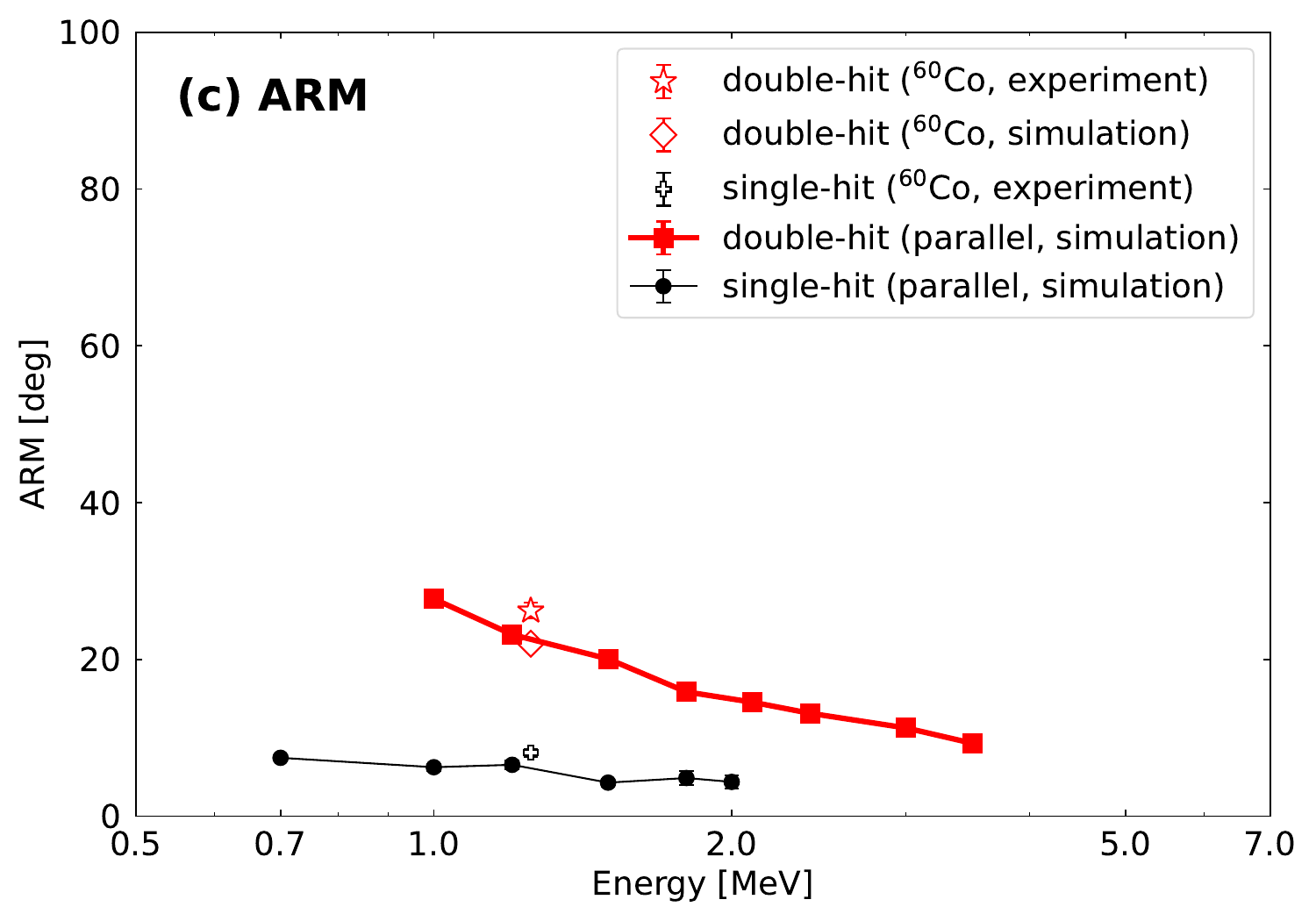}
        \end{minipage}
        &
        \hspace{-0.05\columnwidth}
        \begin{minipage}{0.485\hsize}
            \includegraphics[width=\hsize]{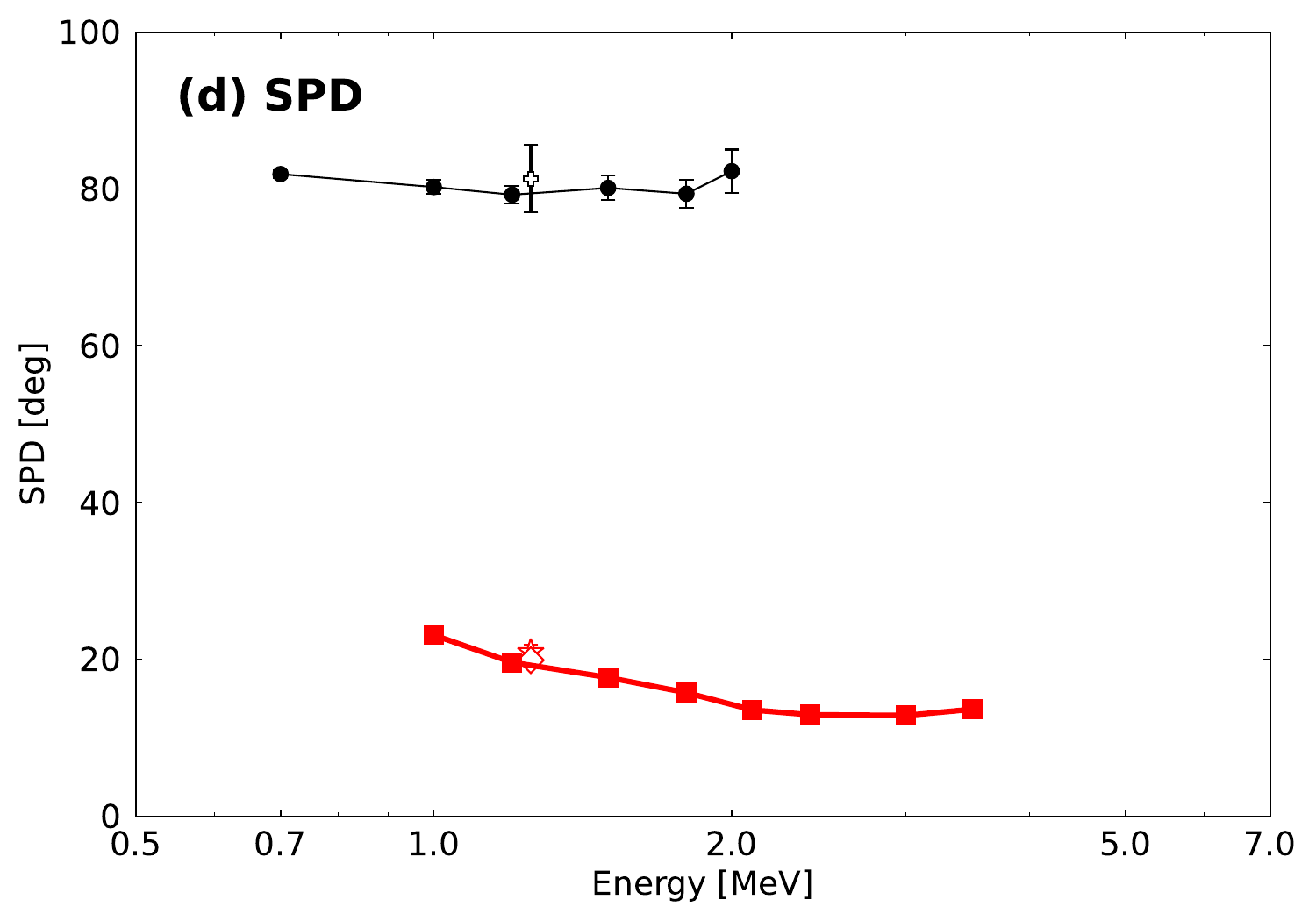}
        \end{minipage}
    \end{tabular}
    \caption{
    The distributions of the (a) ARM and (b) SPD for $^{60}$Co data and (c--d) their energy dependence.
    {\bf (a, b)}  Shaded red  and thick blue-line  histograms  show those of the laboratory data and simulation data, respectively.
    Each histogram is fitted with the Gaussian function,  and the best-fit function is shown  with a solid black line for the laboratory data and  a dashed black line for the simulation data.
    {\bf(c, d)} The ARM (c) and SPD (d) as a function of incident gamma-ray energy.
    Red squares and black circles  denote the results of the parallel-light simulations for double-hit events, whereas black circles  do those of single-hit events.
    Also shown with open markers are the results of $^{60}$Co data.
    Red diamond and star markers  denote the results of double-hit events with the laboratory and simulation data, respectively,  whereas  black  cross markers  do the results of single-hit events with the laboratory data.
    }
    \label{fig:ARM_and_SPD}
\end{figure*}

The energy resolution is found to be $12.8 \pm 0.4\%$ at 1.5~MeV, which is 1.4 times inferior to the single-hit event analysis (Panel a).
This difference in resolution is due to the fact that  two more terms ($E_{\rm{e,\,DZ}}$ and $E_{\rm{e,\,PSA}}$) with some uncertainty in the detector response are considered in the energy reconstruction for double-hit events than for single-hit events (Eq.~\ref{eq:totalenergy}).
Despite the fact that the ARM is affected by this deterioration in energy resolution, the HPR is found to be $8.4^{\circ}$ at 1.5~MeV, which is better than the single-hit event analysis by a factor of 2.0 (Panel b).
This improvement in the HPR is attributed to the significant improvement in the SPD.
Figs.~\ref{fig:ARM_and_SPD}a and \ref{fig:ARM_and_SPD}b show the distributions of the ARM and SPD for the $^{60}$Co data.
Fitting each of these histograms with a Gaussian function\footnote{Generally, the distribution of the ARM should follow a Lorentzian function due to a Doppler broadening~\cite{Zoglauer2003SPIE}. However,  in double-hit events, only the events of relatively high-energy electrons are observed, and thus,  their energy spectra can be approximated  by  a Gaussian function.} finds the resolution of the ARM  to be $26.2 \pm 1.0^{\circ}$ ($22.0 \pm 0.5^{\circ}$) in FWHM for the laboratory (simulation) data, which is worse than that of the single-hit events ($8.1 \pm 0.5^{\circ}$ in FWHM for $^{60}$Co source).
This degradation in the ARM can be explained with the deterioration in the energy resolution.
In the same manner, the resolutions for the SPD  are found to be $19.9 \pm 0.4^{\circ}$ and $20.9 \pm 1.0^{\circ}$ in FWHM for the simulation and laboratory data, respectively, which are significantly better than that ($81.3 \pm 4.3^{\circ}$) of the single-hit events.
This improvement is mainly attributed to the advantage with the geometric estimation of the recoil-electron direction and also a less significant effect of multiple scattering on high-energy recoil electrons than in the single-event analyses.
In addition, panels c and d of Fig.~\ref{fig:ARM_and_SPD} show the energy dependence of the ARM and SPD evaluated with the parallel-light simulations. 
The finding that SPD is significantly improved at the expense of the ARM degradation holds over the entire observation band.

We also find an improvement of effective area in the double-hit event analysis.
The effective area  for single-hit events is limited by the cross section of photoelectric absorption and the probability of confining recoil electrons in the TPC and decreases with energy.
By contrast, the effective area for double-hit events is almost constant at (5.0--$5.5)\times 10^{-2}~\rm{cm^{2}}$ for 1.5--3.0~MeV, indicating that the analysis band in the ETCC observation is extended to the higher energy.

\section{Discussion} \label{sec:discussion}

The analysis results of simulation and real data are in good agreement (Sect.~\ref{sec:lab}),  confirming that the simulator works as expected for the purpose of  evaluating the detector performance.
We have further found that the double-hit event analysis improves the HPR and effective area  from the single-hit event analysis \cite{Takada2022ApJ}.
This section discusses the performance  evaluated with the parallel-light simulations in Sect.~\ref{sec:sim} and the  potentials for further improvement.

\begin{figure}
    \centering
    \includegraphics[width=\hsize]{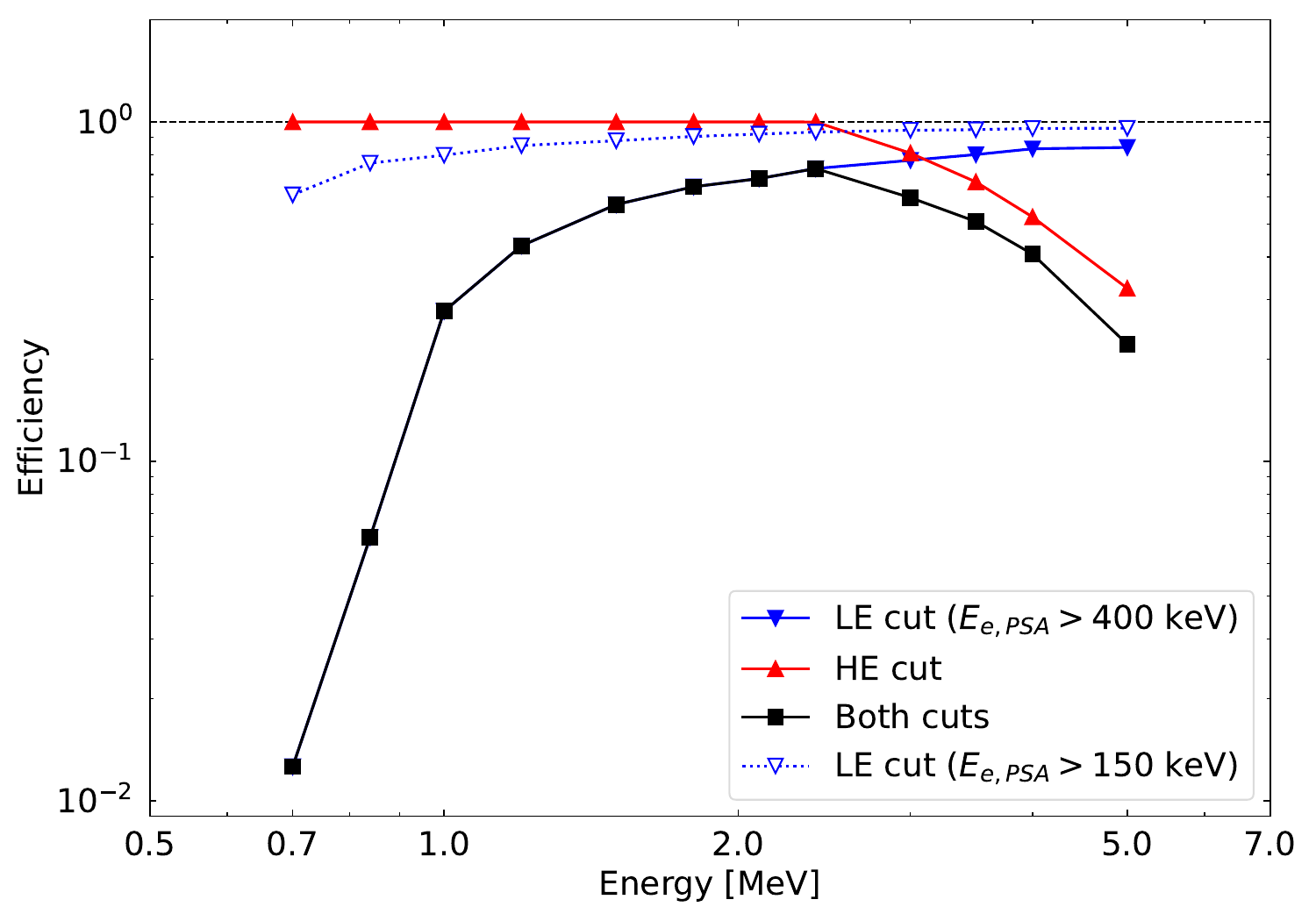}
    \caption{
    Ratio of the remaining data after filtering with energy (i.e., energy cut)  for the simulation data as a function of incident gamma-ray energy.
    Filled red, blue, and black markers show the  results after the high-energy cut (i.e., $E_{\rm PSA} < 2.1$~MeV), the low-energy cut ($E_{\rm{e,\,PSA}} > 400$~keV and $E_{\rm{\gamma,\,PSA}} > 150$~keV), and both cuts, respectively.
    Open blue inverted triangles show the results for the low-energy cut but with $E_{\rm{e,\,PSA}} > 150$~keV (corresponding to the dynamic range of the PSA).
    }
    \label{fig:efficiency_energycut}
\end{figure}

Here, we examine how the energy cut that we have chosen affects the results, including the effective area.
Fig.~\ref{fig:efficiency_energycut} shows the  ratio of the remaining  gamma-ray events  after the application of the energy cut, as a function of incident gamma-ray energy.
The high-energy cut ($E_{\rm PSA} < 2.1$~MeV, defined in Sect.~\ref{sec:rec:cut})  yields a ratio of $52.5\%$ at 4~MeV, whereas the low-energy cut ($E_{\rm e, PSA} > 0.4$~MeV and $E_{\rm \gamma, PSA} > 0.15$~MeV) yields $\lesssim 84.1\%$ at the entire band.
This result implies that these energy cuts significantly affect the effective area of the ETCC.
Since the high-energy cut parameter depends on the dynamic range of the PSAs, adjusting it can result in better efficiency and an increase in the effective area.
The ongoing project, SMILE-3 \cite{Takada2020SPIE}, is improving the ETCC, which includes widening of the dynamic range and thus the energy band (up to ${\gtrsim}10$~MeV).
In that band, the HPR would reach a few degrees  owing to  a smaller multi-scattering effect for higher energy recoil electrons \cite{Bernard2022}.
To further extend the energy band beyond 10~MeV, which is not covered in this paper, one might consider analyzing pair production events instead of Compton scattering events \cite[e.g.,][]{Ueno2011NIMA}.

The low-energy cut is set to accurately estimate the energy loss in the dead zone (Sect.~\ref{sec:lab}).
The performance for low-energy events can be improved by reducing the thickness of the dead zone, which can make  the energy loss  in  the region  negligible, or by using a detector with a small atomic number for the PSAs to reduce further the multiple-scattering effect.
Suppose, for example, that the above-mentioned improvement(s) is  implemented and suppose that the dynamic range of the PSA  defines the low-energy cut as with the high-energy cut.
Then, if the low-energy cut is set at $E_{\rm e\,PSA} > 150$~keV, the  ratio of the remaining events after energy-cut filtering  is improved from $27.7\%$ to $79.7\%$ at 1~MeV, as shown in Fig.~\ref{fig:efficiency_energycut}.
Such a modification for the dead zone may also have a positive effect on the energy resolution and PSF; the exact consequence  would  be  a subject for future experiments.
The modification would also suppress the systematic uncertainty originating in the lack of electron information in the dead zone (see Sect.~\ref{sec:lab}).

\begin{figure}
    \centering
    \begin{tabular}{c}
        \begin{minipage}{\hsize}
            \centering
            \includegraphics[width=\hsize]{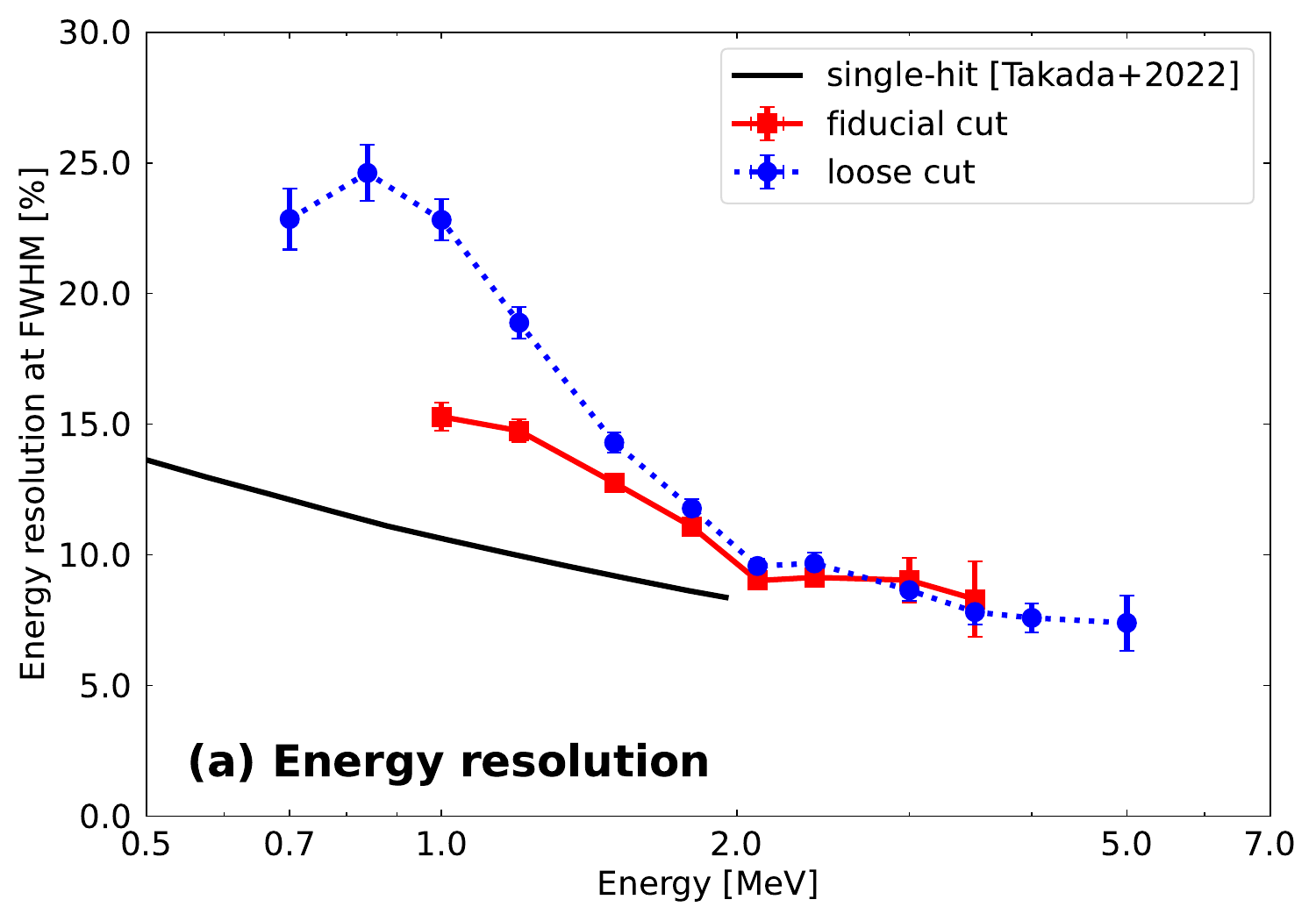}
        \end{minipage}
        \\
        \begin{minipage}{\hsize}
            \centering
            \includegraphics[width=\hsize]{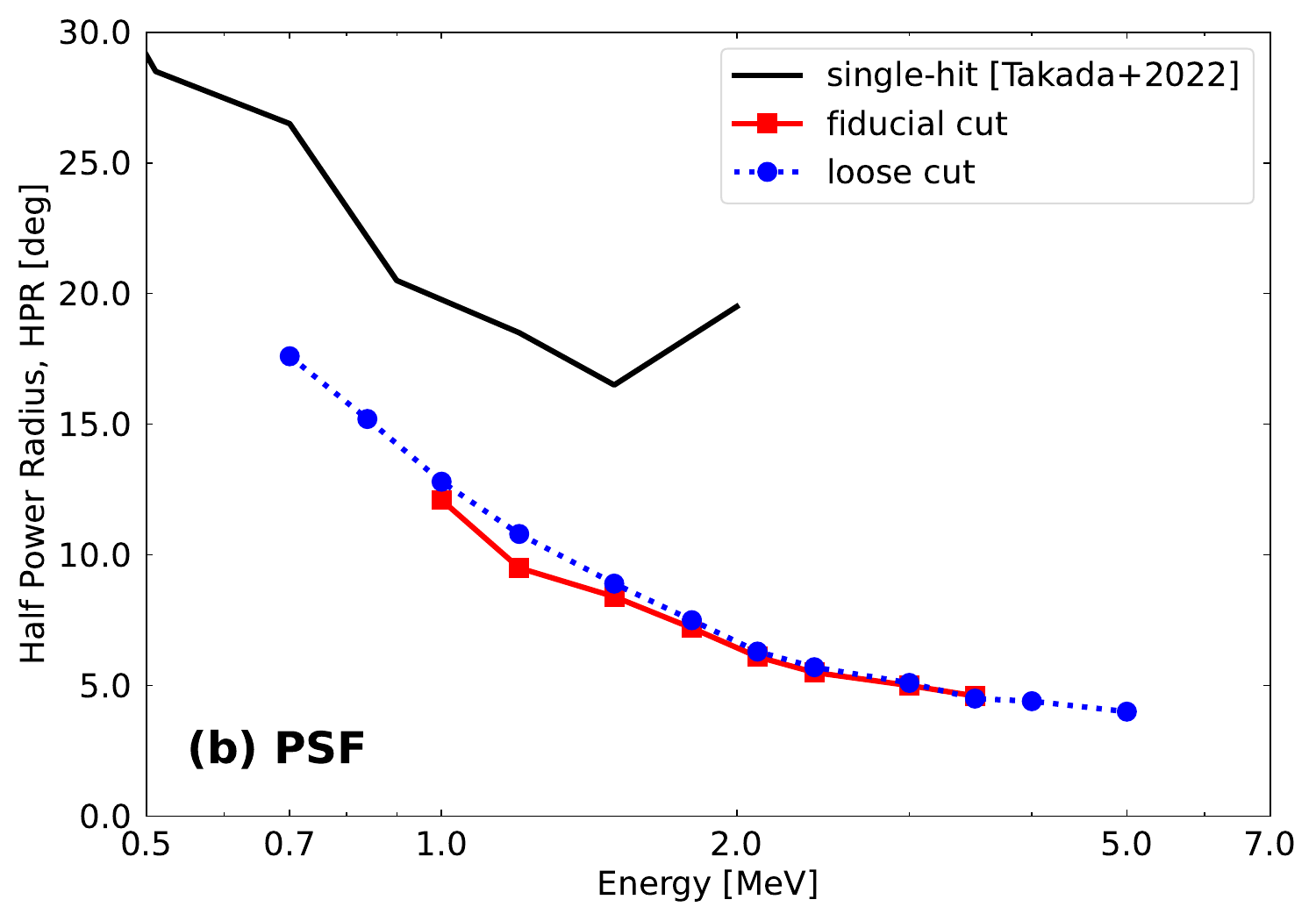}
        \end{minipage}
        \\
        \begin{minipage}{\hsize}
            \centering
            \includegraphics[width=\hsize]{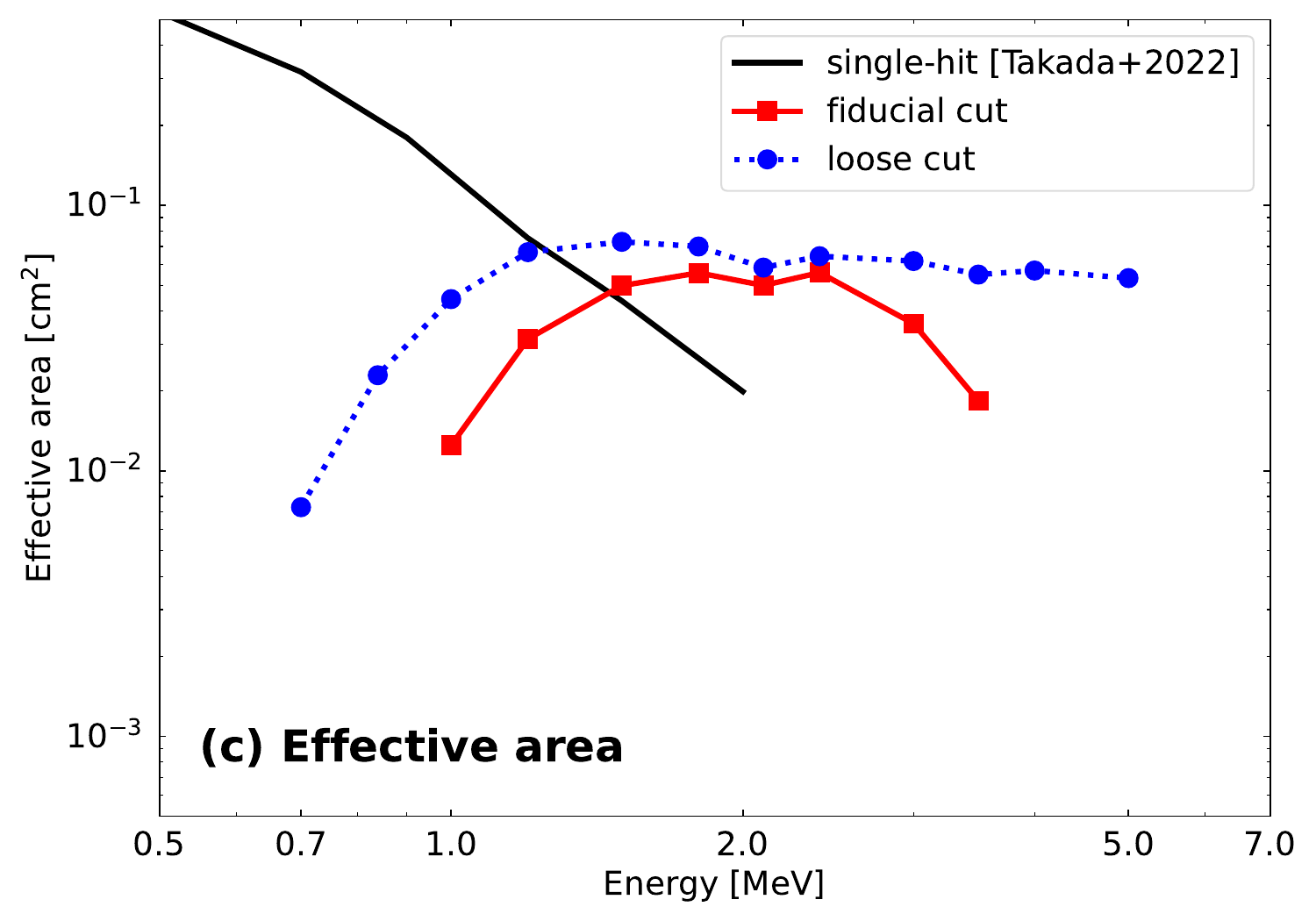}
        \end{minipage}
  \end{tabular}
    \caption{
    Same as Fig.~\ref{fig:performance} but with no energy threshold for the PSAs.
    In each panel, the blue circles show results for the loose cut (i.e., no energy thresholds for the PSAs),  whereas the others are the same as in Fig.~\ref{fig:performance}.
    }
    \label{fig:performance_loosecut}
\end{figure}

Fig.~\ref{fig:performance_loosecut} shows the simulated detector performance without imposing event selection with energy thresholds on PSAs.
The energy resolution at the higher energy band under this loose event selection is comparable with those with the fiducial cut at $>1.5$~MeV, but it is worse at lower energy bands.
This difference at the low energy is due to an increase in the number of events where recoil electrons are not properly absorbed by the scintillator, as shown in Fig.~\ref{fig:DeadZone}(e).
The resultant HPRs with the two cut conditions are in comparable with other and better than $5.0^{\circ}$ at high gamma-ray energies.
The effective area is improved with less stringent cut conditions applied, as expected from Fig.~\ref{fig:efficiency_energycut}, the fact of which suggests that adjusting the PSA dynamic range should improve the sensitivity for higher energy gamma-rays.

\section{Summary} \label{sec:summary}

With the aim of improving the performance of the ETCC for gamma-ray observations above 1~MeV, we have newly developed the analysis method for  double-hit events in the ETCC.
For this type of event, the Compton-recoil electron as well as the scattered gamma ray reach the PSA through the electron tracker.
We have applied this method to the laboratory data and the simulation data.
Comparison of the results between the laboratory and simulation data shows that their energy spectra are in good agreement within a factor of 1.2.
The simulations  also  suggest that the recoil electron may return to the dead zone from the PSA. The fact indicates that  a high energy threshold ($\sim 400~{\rm keV}$) on the TPC is necessary for the analysis to eliminate such events.

We have evaluated the detector performance for double-hit events  with parallel-light simulations.
The energy resolution for 1.5~MeV is $12.8 \pm 0.4\%$  in FWHM,  whereas the angular resolution at the energy is $8.4^{\circ}$ in HPR.
The latter is twice as good as that of the previous study \citep{Takada2022ApJ}.
The effective area is also improved above $1.5~\rm{MeV}$ and is twice as good at 2~MeV as that in the previous works.
The changes in the angular resolution and effective area lead to an improvement in detection sensitivity \cite{Bernard2022}.
A finer-quality study will be conducted in near future with observation data from the SMILE-2+ balloon experiment.
The most important finding in this work is that implementing a double-hit event analysis extends the observable energy band to a higher energy.
Although the maximum energy is currently limited to 3.5~MeV, to stretch it further to a higher energy would be possible using PSAs with a larger dynamic range, which is already being adopted in the ongoing project, SMILE-3.
Accordingly, the ETCC could target the $^{26}$Al line emission at 1.8~MeV, the nuclear de-excitation line emissions from molecular clouds at 4--6~MeV, and the COMPTEL excess at ${\gtrsim}1$~MeV as listed in Sect.~\ref{sec:introduction}.
Application of the double-hit event analysis makes it feasible to achieve a better sensitivity in the MeV gamma-ray band, leading to obtaining the most precise MeV gamma-ray sky map in future observations with the ETCC.

\section*{Acknowledgments}

We would like to thank Junko Kushida, Masaki Mori, and Takeshi Nakamori for useful discussion.
We also thank the anonymous referees for valuable comments.
This study was supported by the Japan Society for the Promotion of Science (JSPS) Grant-in-Aid for Scientific Research (S) (21224005), (A) (20244026, 16H02185), Grant-in-Aid for Young Scientists\ (B) (15K17608), JSPS Grant-in-Aid for Challenging Exploratory Research (23654067, 25610042, 16K13785, 20K20428), a Grant-in-Aid from the Global COE program ``Next Generation Physics, Spun from Universality and Emergence'' from the Ministry of Education, Culture, Sports, Science and Technology (MEXT) of Japan, Early-Career Scientists (22K14057), and Grant-in-Aid for JSPS Fellows (16J08498, 18J20107, 19J11323, 22KJ1766, 23KJ2094). 
Some of the electronics development was supported by KEK-DTP and Open-It Consortium.

\bibliography{my_bib}

\end{document}